\documentclass[aps,prd,twocolumn,superscriptaddress]{revtex4}
\usepackage{epsfig,epsf}
\usepackage{amsmath}
\usepackage{amsthm}
\usepackage{amsfonts}
\usepackage{amssymb}
\usepackage{dsfont}
\usepackage{multirow}
\usepackage{appendix}
\usepackage{slashed}
\usepackage[active]{srcltx}
\usepackage{psfrag}

\begin{document}

\title{A family of double-beauty tetraquarks: Axial-vector state $T_{bb;%
\overline{u}\overline{s}}^{-}$ }
\author{S.~S.~Agaev}
\affiliation{Institute for Physical Problems, Baku State University, Az--1148 Baku,
Azerbaijan}
\author{K.~Azizi}
\affiliation{Department of Physics, University of Tehran, North Karegar Avenue, Tehran
14395-547, Iran}
\affiliation{Department of Physics, Do\v{g}u\c{s} University, Acibadem-Kadik\"{o}y, 34722
Istanbul, Turkey}
\author{B.~Barsbay}
\affiliation{Department of Physics, Do\v{g}u\c{s} University, Acibadem-Kadik\"{o}y, 34722
Istanbul, Turkey}
\affiliation{Department of Physics, Kocaeli University, 41380 Izmit, Turkey}
\author{H.~Sundu}
\affiliation{Department of Physics, Kocaeli University, 41380 Izmit, Turkey}

\begin{abstract}
The spectroscopic parameters and decay channels of the axial-vector
tetraquark $T_{bb;\overline{u}\overline{s}}^{-}$ (in what follows, $T_{b:%
\overline{s}}^{\mathrm{AV}}$) are explored using the quantum chromodynamics (QCD) sum rule
method. The mass and coupling of this state are calculated using 
two-point sum rules by taking into account various vacuum condensates, up to
10 dimensions. Our prediction for the mass of this state $m=(10215\pm 250)~%
\mathrm{MeV}$ confirms that it is stable with respect to strong and electromagnetic
decays and can dissociate to conventional mesons only via weak
transformations. We investigate the dominant semileptonic $T_{b:\overline{s}%
}^{\mathrm{AV}} \to \mathcal{Z}_{b:\overline{s}}^{0}l\overline{\nu}_l$ and
nonleptonic $T_{b:\overline{s}}^{\mathrm{AV}} \to \mathcal{Z}_{b:\overline{s}%
}^{0}M$ decays of $T_{b:\overline{s}}^{\mathrm{AV}}$. In these processes, $%
\mathcal{Z}_{b:\overline{s}}^{0}$ is a scalar tetraquark $[bc][\overline{u}%
\overline{s}]$ built of a color-triplet diquark and an antidiquark, whereas $M$
is one of the vector mesons $\rho ^{-}$, $K^{\ast}(892)$, $D^{\ast
}(2010)^{-}$, and $D_{s}^{\ast -}$. To calculate the partial widths of
these decays, we  use  the QCD three-point sum rule approach and
evaluate the weak transition form factors $G_{i}$ $(i=0,1,2,3)$, which govern
these processes. The full width $\Gamma _{\mathrm{full}} =(12.9\pm
2.1)\times 10^{-8}~\mathrm{MeV}$ and the mean lifetime $%
\tau=5.1_{-0.71}^{+0.99}~\mathrm{fs}$ of the tetraquark $T_{b:\overline{s}}^{%
\mathrm{AV}}$ are computed using the aforementioned weak decays. The obtained
information about the  parameters of $T_{b:\overline{s}}^{\mathrm{AV}}$ and $%
\mathcal{Z}_{b:\overline{s}}^{0}$ is useful for experimental investigations
of these double-heavy exotic mesons.
\end{abstract}

\maketitle


\section{Introduction}

\label{sec:Int}
Recently, double-beauty tetraquarks, composed of a $bb$ diquark and a light
antidiquark $\overline{q}\overline{q}^{\prime }$, became a subject of
intensive theoretical studies \cite%
{Karliner:2017qjm,Eichten:2017ffp,Agaev:2018khe,Hernandez:2019eox,Ali:2018ifm,Ali:2018xfq}%
. The interest in these states was inspired by the experimental observation of
baryons $\Xi _{cc}^{++}$ and measurements of their parameters \cite%
{Aaij:2017ueg}. The measurements were used in
phenomenological models, to estimate the masses of double-beauty states \cite%
{Karliner:2017qjm}. These investigations demonstrated that
the axial-vector tetraquark $T_{bb;\overline{u}\overline{d}}^{-}$ (hereafter $%
T_{bb}^{-}$) with the mass $m=(10389\pm 12)~\mathrm{MeV}$ is stable with respect to
strong and electromagnetic decays and can dissociate into a conventional meson
only via a weak transformation. A similar conclusion about the stable nature
of some tetraquarks $bb\overline{q}\overline{q}^{\prime }$ was reached in Ref.\
\cite{Eichten:2017ffp} as well, where the authors of that study used methods of
heavy-quark symmetry analysis.

Double-heavy tetraquarks $QQ^{\prime }\overline{q}\overline{q}^{\prime }$%
, in fairness, were studied already in classical articles \cite%
{Ader:1981db,Lipkin:1986dw,Zouzou:1986qh,Carlson:1987hh,Manohar:1992nd}, in
which they were examined as candidate stable four-quark compounds. The
main qualitative conclusion drawn in these works was the existence of a constraint on the masses of
constituent quarks. It was found that tetraquarks $QQ^{\prime }\overline{q}%
\overline{q}^{\prime }$ may form strong-interaction stable exotic mesons,
provided the ratio $m_{Q}/m_{q}$ is large. Therefore, tetraquarks $bb%
\overline{q}\overline{q}^{\prime }$ are the most promising candidates for stable
four-quark mesons.

Quantitative analysis of these problems continued in the following years, using the 
frameworks of various models and using different methods from high-energy
physics. Thus, tetraquarks $T_{QQ}$ were explored using the chiral,
dynamical, and relativistic quark models \cite%
{Pepin:1996id,Janc:2004qn,Cui:2006mp,Vijande:2006jf,Ebert:2007rn}.
Axial-vector states $T_{QQ;\overline{u}\overline{d}}$ were considered in the
context of the sum rule method \cite{Navarra:2007yw,Du:2012wp}. Processes in
which tetraquarks $T_{cc}$ may be produced, namely electron-positron
annihilation, heavy-ion and proton-proton collisions, and $B_{c}$ meson and $%
\Xi _{bc}$ baryon decays, also attracted the interest of researchers \cite%
{SchaffnerBielich:1998ci,DelFabbro:2004ta,Lee:2007tn,Hyodo:2012pm,Esposito:2013fma}%
.

The axial-vector particle $T_{bb}^{-}$ was studied in our work as well \cite%
{Agaev:2018khe}. We employed the quantum chromodynamics (QCD) sum rule method and evaluated the mass
of this state $m=(10035~\pm 260)~\mathrm{MeV}$. This means that $m$ is below
both the $B^{-}\overline{B}^{\ast 0}$ and $B^{-}\overline{B}^{0}\gamma $
thresholds; hence, this state is a strong- and electromagnetic-interaction
stable tetraquark. We also explored the semileptonic decays $T_{bb}^{-}$ $%
\rightarrow Z_{bc}^{0}l\overline{\nu }_{l}$, where $Z_{bc}^{0}$ is the
scalar tetraquark $[bc][\overline{u}\overline{d}]$ composed of color-triplet
diquarks, and calculated their partial widths. The predictions for the full
width and mean lifetime of $T_{bb}^{-}$ obtained in Ref.\ \cite%
{Agaev:2018khe} are useful for experimental investigations of double-beauty
exotic mesons.

Other members of the $bb\overline{q}\overline{q}^{\prime }$ family, studied
in a rather detailed form, are the scalar tetraquarks $T_{bb;\overline{u}%
\overline{s}}^{-}$ and $T_{bb;\overline{u}\overline{d}}^{-}$ (in short
forms, $T_{b:\overline{s}}^{-}$ and $T_{b:\overline{d}}^{-}$, respectively).
The mass and coupling of $T_{b:\overline{s}}^{-}$ and $T_{b:\overline{d}%
}^{-} $ were calculated in Refs.\ \cite{Agaev:2019lwh,Agaev:2020dba}, in
which we demonstrated that they cannot decay to ordinary mesons through
strong and electromagnetic processes. We also investigated dominant
semileptonic and nonleptonic weak decays of these tetraquarks and estimated
their full width and lifetime characteristics.

In the present article, we extend our analysis and investigate the
axial-vector partner of $T_{b:\overline{s}}^{-}$ with the same quark
content $bb\overline{u}\overline{s}$. It can be treated also as "$s$" member
of the axial-vector multiplet of the states $bb\overline{u}\overline{q}$. We
denote this tetraquark as $T_{b:\overline{s}}^{\mathrm{AV}}$ and compute its
spectroscopic parameters using the two-point QCD sum rule method.
Calculations are performed by taking into account various vacuum
condensates, up to 10 dimensions. The obtained result for its mass $%
m=(10215\pm 250)~\mathrm{MeV}$ proves that this state is stable against
strong and electromagnetic decays. In fact, $T_{b:\overline{s}}^{\mathrm{AV}%
} $ in the $S$-wave can decompose into pairs of conventional mesons $%
B^{-}B_{s}^{\ast }$ and $B^{\ast -}\overline{B}_{s}^{0}$, provided $m$
exceeds the corresponding thresholds $10695/10692~\mathrm{MeV}$,
respectively. The threshold for the electromagnetic decay to the final state $B^{-}\overline{%
B}_{s}^{0}\gamma $ is $10646~\mathrm{MeV}$. It is seen
that even the maximal allowed value of the mass $10465~\mathrm{MeV}$ is
below all of these limits.

Therefore, to evaluate the full width and lifetime of $T_{b:\overline{s}}^{%
\mathrm{AV}}$, we analyzed the semileptonic and nonleptonic weak decays $T_{b:%
\overline{s}}^{\mathrm{AV}}\rightarrow \mathcal{Z}_{b:\overline{s}}^{0}l%
\overline{\nu }_{l}$ and $T_{b:\overline{s}}^{\mathrm{AV}}\rightarrow
\mathcal{Z}_{b:\overline{s}}^{0}M$, respectively. Here, $\mathcal{Z}_{b:%
\overline{s}}^{0}$ is the scalar tetraquark $[bc][\overline{u}\overline{s}]$
built of a color-triplet diquark and an antidiquark, and $M$ is one of the
vector mesons $\rho ^{-}$, $K^{\ast }(892)$, $\ D^{\ast }(2010)^{-}$, and $\
D_{s}^{\ast -}$. The weak transitions of $T_{b:\overline{s}}^{\mathrm{AV}}$
can be described by the form factors $G_{i}(q^{2})$ \ ($i=0,1,2,3$),
which determine the differential rates $d\Gamma /dq^{2}$ of the semileptonic and
partial widths of the nonleptonic processes. These weak form factors are
extracted from the QCD three-point sum rules in Section \ref{sec:Decays1}.

This work is structured as follows. In Section \ref{sec:Masses},
we calculate the mass and coupling of the tetraquarks $T_{b:\overline{s}}^{%
\mathrm{AV}}$ and $\mathcal{Z}_{b:\overline{s}}^{0}$. For this, we
derive the sum rules for their masses and couplings, by analyzing the
corresponding two-point correlation functions. Numerical computations are
performed by taking into account quark, gluon, and mixed condensates, up to
the 10th dimension. In Section \ref{sec:Decays1}, we compute the weak form
factors $G_{i}(q^{2})$ from the three-point sum rules for momentum transfers $%
q^{2}$, where this method is applicable. In that section, we also determine
model functions $\mathcal{G}_{i}(q^{2})$ and find the partial widths of the
semileptonic decays\ $T_{b:\overline{s}}^{\mathrm{AV}}\rightarrow \mathcal{Z}%
_{b:\overline{s}}^{0}l\overline{\nu }_{l}$ . The weak nonleptonic processes $%
T_{b:\overline{s}}^{\mathrm{AV}}\rightarrow \mathcal{Z}_{b:\overline{s}%
}^{0}M $ are investigated in Section \ref{sec:Decays2}. This section
also contains our final results for the full width and mean lifetime of the
tetraquark $T_{b:\overline{s}}^{\mathrm{AV}}$. In Section \ref{sec:Disc} we
discuss our obtained results and present our conclusions. Appendix
contains explicit expressions of quark propagators and the correlation
function used to evaluate the parameters of the tetraquark $T_{b:\overline{s}}^{%
\mathrm{AV}}$.


\section{Spectroscopic parameters of the axial-vector $T_{b:\overline{s}}^{%
\mathrm{AV}}$ and scalar $\mathcal{Z}_{b:\overline{s}}^{0}$ tetraquarks}

\label{sec:Masses}
In this section, we calculate the mass $m_{\mathrm{AV}}$ and coupling $f_{%
\mathrm{AV}}$ of the axial-vector tetraquark $T_{b:\overline{s}}^{\mathrm{AV}%
}$, which is necessary for clarifying its nature, and conclude whether this
particle is stable against strong and electromagnetic decays.
Another tetraquark considered here is the scalar exotic meson $\mathcal{Z}%
_{b:\overline{s}}^{0}$ that appears in the final state of the master
particle's decays: spectroscopic parameters of this state enter into
the expressions for the partial widths of the $T_{b:\overline{s}}^{\mathrm{AV}}$
tetraquark's decay channels. The scalar exotic meson $\mathcal{Z}_{b:\overline{s}}^{0}$ is a
member of the $bc\overline{q}\overline{q}^{\prime }$ family and is of
interest from this perspective as well.

The sum rules for evaluating the mass and coupling of the axial-vector
tetraquark $T_{b:\overline{s}}^{\mathrm{AV}}$ can be obtained from 
the two-point correlation function
\begin{equation}
\Pi _{\mu \nu }(p)=i\int d^{4}xe^{ipx}\langle 0|\mathcal{T}\{J_{\mu
}(x)J_{\nu }^{\dag }(0)\}|0\rangle ,  \label{eq:CF1vector}
\end{equation}%
where $J_{\mu }(x)$ is the corresponding interpolating current.  It is known
that there are five independent diquark fields without derivatives, which
can be used for formulating the current $J_{\mu }(x)$. Among them, scalar and
axial-vector diquarks are the most stable and favorable structures for composing the
tetraquark state. We suggest that $T_{b:\overline{s}}^{\mathrm{AV}}$ is
composed of the axial-vector diquark $b^{T}C\gamma _{\mu }b$ and the scalar
antidiquark $\overline{u}\gamma _{5}C\overline{s}^{T}$. One has to take into
account that the axial-vector diquark $b^{T}C\gamma _{\mu }b$ has symmetric
flavor but antisymmetric color organization, and its flavor-color structure
is fixed as $(\mathbf{6}_{f},\overline{\mathbf{3}}_{c})$ \cite{Du:2012wp}.
Then, to build a color-singlet current, the light antidiquark field should belong
to the triplet representation of the $SU_{c}(3)$ color group and has the explicit
form $\overline{u}_{a}\gamma _{5}C\overline{s}_{b}^{T}-\overline{u}%
_{b}\gamma _{5}C\overline{s}_{a}^{T}$. But in calculations, owing to the
symmetry constraint, it is sufficient to keep one of the light diquark terms \cite%
{Du:2012wp}.  Therefore, for the current\ $J_{\mu }(x)$ we use the
following expression
\begin{equation}
J_{\mu }(x)=\left[ b_{a}^{T}(x)C\gamma _{\mu }b_{b}(x)\right] \left[
\overline{u}_{a}(x)\gamma _{5}C\overline{s}_{b}^{T}(x)\right] .
\label{eq:CurrVector}
\end{equation}%
To solve the same problems in the case of the scalar tetraquark $\mathcal{Z}%
_{b:\overline{s}}^{0}$, we start from the correlation function
\begin{equation}
\Pi (p)=i\int d^{4}xe^{ipx}\langle 0|\mathcal{T}\{J_{\mathcal{Z}}(x)J_{%
\mathcal{Z}}^{\dag }(0)\}|0\rangle .  \label{eq:CF1}
\end{equation}%
Here, $J_{\mathcal{Z}}(x)$ is the interpolating current for $\mathcal{Z}_{b:%
\overline{s}}^{0}$%
\begin{eqnarray}
J_{\mathcal{Z}}(x) &=&[b_{a}^{T}(x)C\gamma _{5}c_{b}(x)]\left[ \overline{u}%
_{a}(x)\gamma _{5}C\overline{s}_{b}^{T}(x)\right.   \notag \\
&&-\overline{u}_{b}(x)\gamma _{5}C\overline{s}_{a}^{T}(x)].
\label{eq:CurrScalar}
\end{eqnarray}%
In the expressions above, $a$ and $b$ are the color indices, and $C$ is the charge
conjugation operator. The current (\ref{eq:CurrScalar}) is composed of
diquarks that belong to the triplet representation $[\overline{\mathbf{3}}%
_{c}]_{bc}\otimes \lbrack \mathbf{3}_{c}]_{\overline{u}\overline{s}}$ of the
color group.

Now, we concentrate on calculating the parameters $m_{\mathrm{AV}}$ and $%
f_{\mathrm{AV}}$. Following the standard prescriptions of the sum rule method,
we express $\Pi _{\mu \nu }(p)$ using the spectroscopic parameters of $T_{b:%
\overline{s}}^{\mathrm{AV}}$. These manipulations generate the physical or
phenomenological side of the sum rules $\Pi _{\mu \nu }^{\mathrm{Phys}}(p)$
\begin{equation}
\Pi _{\mu \nu }^{\mathrm{Phys}}(p)=\frac{\langle 0|J_{\mu }|T_{b:\overline{s}%
}^{\mathrm{AV}}(p)\rangle \langle T_{b:\overline{s}}^{\mathrm{AV}}(p)|J_{\nu
}^{\dagger }|0\rangle }{m_{\mathrm{AV}}^{2}-p^{2}}+\cdots.
\label{eq:CF2vector}
\end{equation}%
Here, we isolate the ground-state contribution to $\Pi _{\mu \nu }^{\mathrm{%
Phys}}(p)$ from the effects due to higher resonances and continuum states, which
are denoted by dots. In our study, we assume that the phenomenological
side of the sum rules $\Pi _{\mu \nu }^{\mathrm{Phys}}(p)$ can be
approximated by a zero-width single-pole term. In the case of the four-quark
system, the physical side, however, also contains contributions from
two-meson reducible terms \cite{Kondo:2004cr,Lee:2004xk}. Interaction of $%
J_{\mu }(x)$ with such a two-meson continuum generates a finite width $\Gamma
(p^{2})$ of the tetraquark and results in the following modification \cite%
{Wang:2015nwa}:
\begin{equation}
\frac{1}{m^{2}-p^{2}}\rightarrow \frac{1}{m^{2}-p^{2}-i\sqrt{p^{2}}\Gamma
(p^{2})}.  \label{eq:Modification}
\end{equation}%
The contribution of the two-meson continuum can be properly taken into account
by rescaling the coupling $f$, whereas the mass of the tetraquark $m$
preserves its initial value \cite{Sundu:2018nxt}. These effects may be
essential for strong-interaction unstable tetraquarks, because their full
widths are a few $100~\mathrm{MeV}$. Stated differently, the
two-meson continuum is important, provided the mass of the tetraquark is higher
than a relevant threshold. However, even in the case of unstable tetraquarks,
these effects are numerically small; therefore, it is convenient for the
phenomenological side to use Eq.\ (\ref{eq:CF2vector}) and perform an a posteriori 
self-consistency check of obtained results by estimating two-meson contributions
\cite{Sundu:2018nxt}. As we shall see later, the tetraquark $T_{b:\overline{s%
}}^{\mathrm{AV}}$ is a strong-interaction stable particle, and $m_{\mathrm{AV}}$
resides below the two-meson continuum, which justifies the zero-width
single-pole approximation for $\Pi _{\mu \nu }^{\mathrm{Phys}}(p).$

The correlator $\Pi _{\mu \nu }^{\mathrm{Phys}}(p)$ can be simplified
further by defining the matrix element $\langle 0|J_{\mu }|T_{b:\overline{s}%
}^{\mathrm{AV}}(p)\rangle $
\begin{equation}
\langle 0|J_{\mu }|T_{b:\overline{s}}^{\mathrm{AV}}(p)\rangle =m_{\mathrm{AV}%
}f_{\mathrm{AV}}\epsilon _{\mu },  \label{eq:MElem1}
\end{equation}%
where $\epsilon _{\mu }$ is the polarization vector of the state $T_{b:%
\overline{s}}^{\mathrm{AV}}$. In terms of $m_{\mathrm{AV}}$ and $f_{\mathrm{%
AV}}$\ , the function $\Pi _{\mu \nu }^{\mathrm{Phys}}(p)$ takes the form
\begin{equation}
\Pi _{\mu \nu }^{\mathrm{Phys}}(p)=\frac{m_{\mathrm{AV}}^{2}f_{\mathrm{AV}%
}^{2}}{m_{\mathrm{AV}}^{2}-p^{2}}\left( -g_{\mu \nu }+\frac{p_{\mu }p_{\nu }%
}{m_{\mathrm{AV}}^{2}}\right) +\cdots.  \label{eq:CorM}
\end{equation}

The QCD side of the sum rules can be found by substituting $J_{\mu }(x)$
into the correlation function (\ref{eq:CF1vector}) and contracting the
relevant quark fields, which yields
\begin{eqnarray}
&&\Pi _{\mu \nu }^{\mathrm{OPE}}(p)=i\int d^{4}xe^{ipx}\mathrm{Tr}\left[
\gamma _{5}\widetilde{S}_{s}^{b^{\prime }b}(-x)\gamma _{5}S_{u}^{a^{\prime
}a}(-x)\right]  \notag \\
&&\times \left\{ \mathrm{Tr}\left[ \gamma _{\nu }\widetilde{S}%
_{b}^{ba^{\prime }}(x)\gamma _{\mu }S_{b}^{ab^{\prime }}(x)\right] -\mathrm{%
Tr}\left[ \gamma _{\nu }\widetilde{S}_{b}^{aa^{\prime }}(x)\gamma _{\mu
}S_{b}^{bb^{\prime }}(x)\right] \right\},  \notag \\
&&  \label{eq:CF3vector}
\end{eqnarray}%
where $S_{q}^{ab}(x)$ is the quark propagator. The propagators of heavy and
light quarks used in the present work are presented in Appendix. In Eq.\ (%
\ref{eq:CF3vector}), we introduce the notation
\begin{equation}
\widetilde{S}_{q}(x)=CS_{q}^{T}(x)C.  \label{eq:Prop}
\end{equation}

It is seen that the correlator $\Pi _{\mu \nu }^{\mathrm{Phys}}(p)$ contains
the Lorentz structure of the vector particle. To derive the sum rules, we
choose to work with invariant amplitudes $\Pi ^{\mathrm{Phys}}(p^{2})$ and $%
\Pi ^{\mathrm{OPE}}(p^{2})$ corresponding to terms $\sim g_{\mu \nu }$,
because they are free of the scalar particles' contributions.

\begin{widetext}

\begin{figure}[h!]
\begin{center}
\includegraphics[totalheight=6cm,width=8cm]{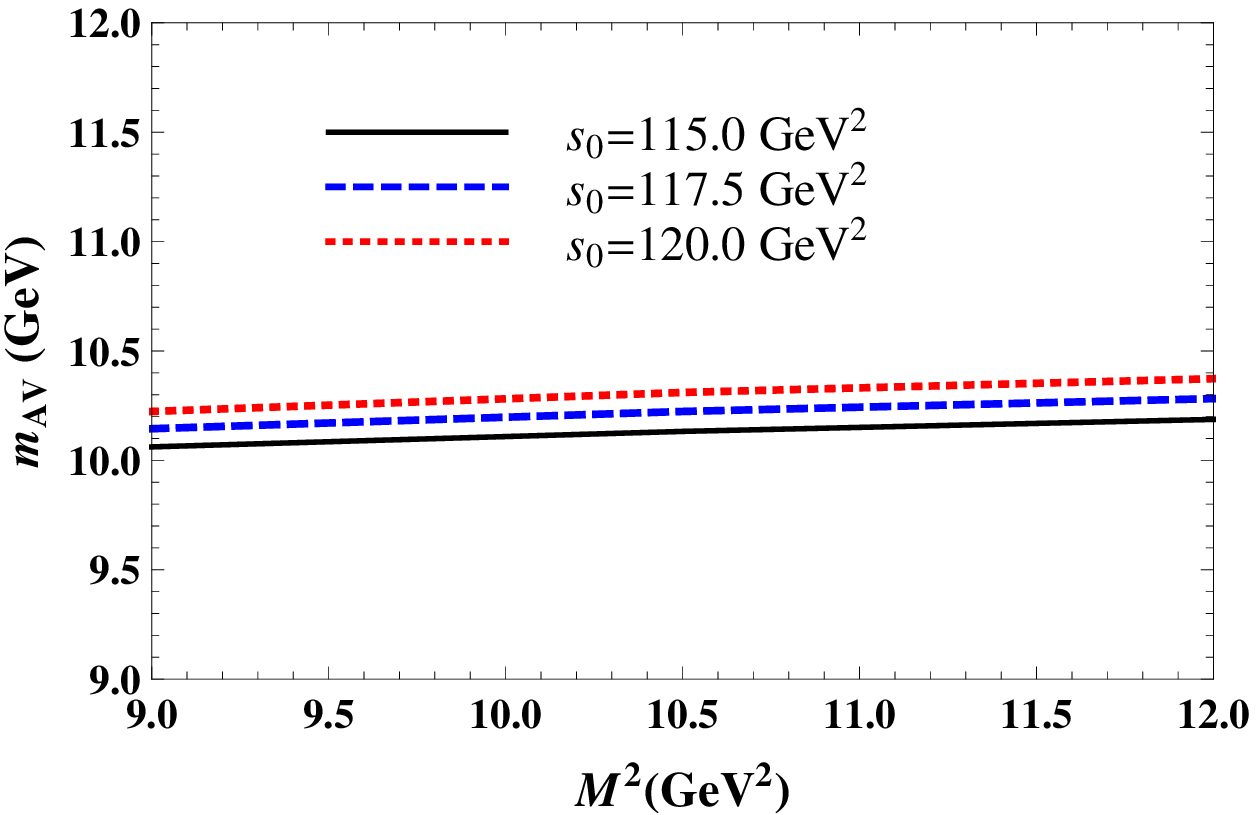}\,\, %
\includegraphics[totalheight=6cm,width=8cm]{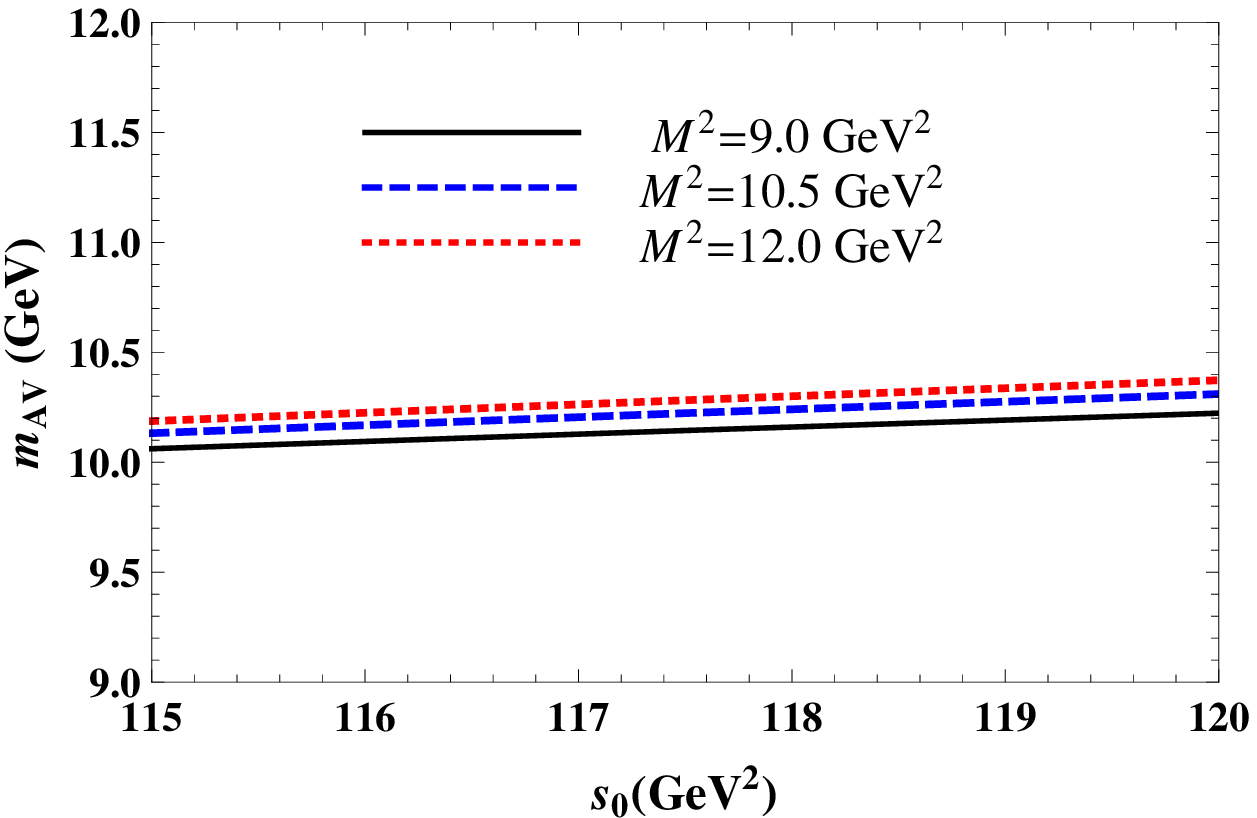}
\end{center}
\caption{  Dependence of the mass $m_{\mathrm{AV}}$ on the Borel $M^{2}$ (left panel) and
continuum threshold $s_{0}$ parameters (right panel).}
\label{fig:MassVector}
\end{figure}

\begin{figure}[h!]
\begin{center}
\includegraphics[totalheight=6cm,width=8cm]{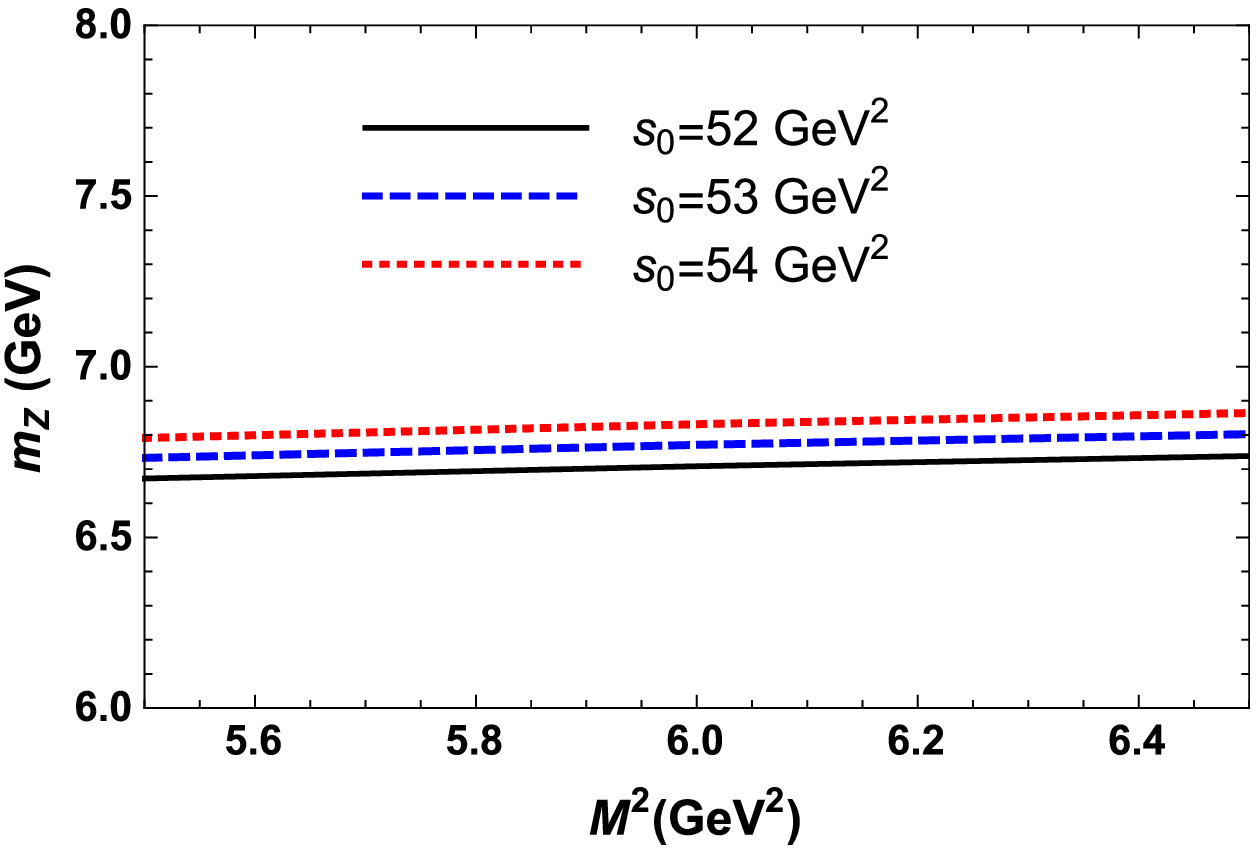}\,\, %
\includegraphics[totalheight=6cm,width=8cm]{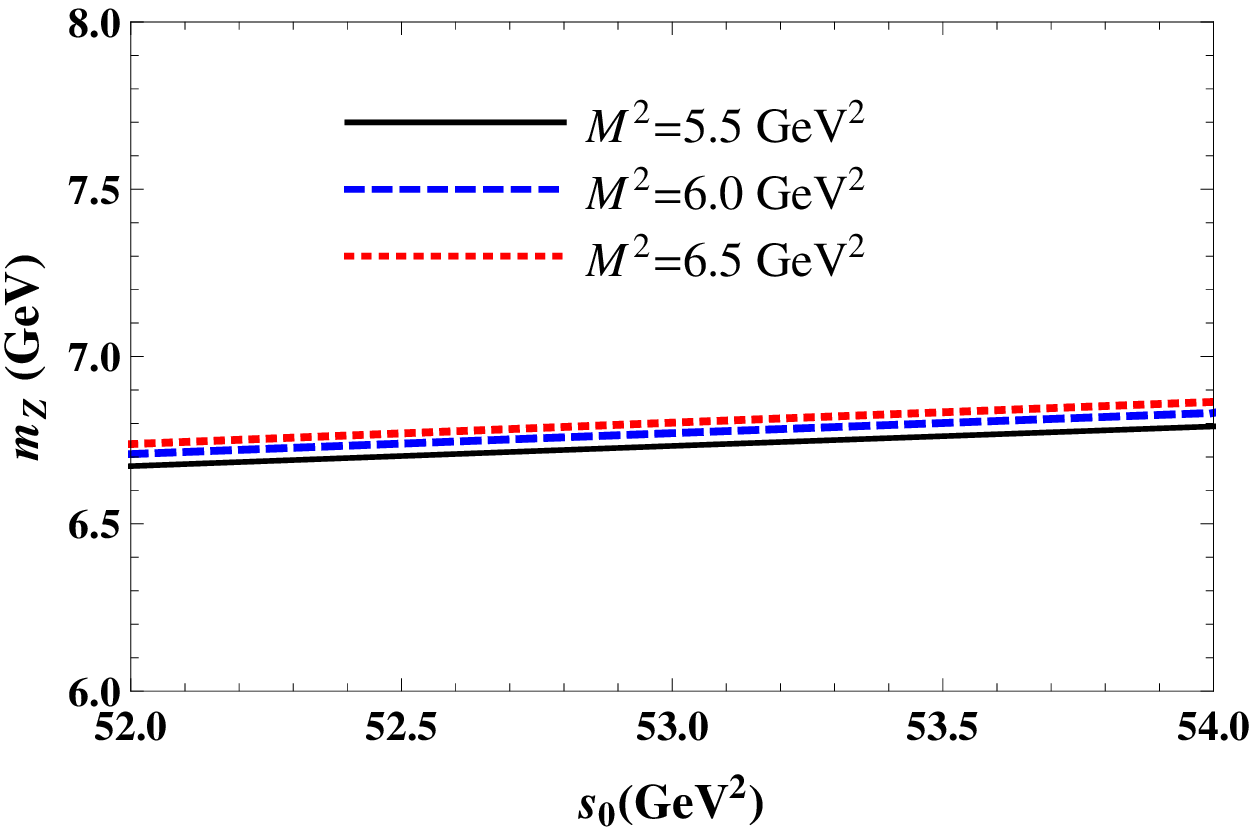}
\end{center}
\caption{  The mass $m_{\mathcal{Z}}$ of the tetraquark $\mathcal{Z}_{b:%
\overline{s}}^{0}$ as a function of the parameters $M^{2}$ (left panel) and $s_{0}$ (right panel).}
\label{fig:MassZ}
\end{figure}
\end{widetext}

The sum rules for $m_{\mathrm{AV}}$ and $f_{\mathrm{AV}}$ can be derived by
equating these two invariant amplitudes and carrying out all standard
manipulations of the method. In the first stage, we apply the Borel
transformation to the both sides of this equality, which suppresses
the contributions of higher resonances and continuum states. In the next step,
using the quark-hadron duality hypothesis, we subtract the higher resonance
and continuum terms from the physical side of the equality. As a result, the
sum rule equality becomes dependent on the Borel $M^{2}$ and continuum
threshold $s_{0}$ parameters. The second equality necessary for deriving
the required sum rules is obtained by applying the operator $d/d(-1/M^{2})$ to
the first expression. Then, the sum rules for $m_{\mathrm{AV}}$ and $f_{%
\mathrm{AV}}$ are
\begin{equation}
m_{\mathrm{AV}}^{2}=\frac{\Pi ^{\prime }(M^{2},s_{0})}{\Pi (M^{2},s_{0})},
\label{eq:MassVector}
\end{equation}%
and
\begin{equation}
f_{\mathrm{AV}}^{2}=\frac{e^{m_{\mathrm{AV}}^{2}/M^{2}}}{m_{\mathrm{AV}}^{2}}%
\Pi (M^{2},s_{0}).  \label{eq:CouplingV}
\end{equation}%
Here, $\Pi (M^{2},s_{0})$ is the Borel-transformed and continuum-subtracted
invariant amplitude $\Pi ^{\mathrm{OPE}}(p^{2})$, and $\Pi ^{\prime
}(M^{2},s_{0})=d/d(-1/M^{2})\Pi (M^{2},s_{0})$. The function $\Pi
(M^{2},s_{0})$ has the following form:%
\begin{equation}
\Pi (M^{2},s_{0})=\int_{\mathcal{M}^{2}}^{s_{0}}ds\rho ^{\mathrm{OPE}%
}(s)e^{-s/M^{2}}+\Pi (M^{2}),  \label{eq:InvAmp}
\end{equation}%
where $\mathcal{M}=2m_{b}+m_{s}$. The quantity $\rho ^{\mathrm{OPE}}(s)$ is the
two-point spectral density, whereas the second component of the invariant
amplitude $\Pi (M^{2})$ includes nonperturbative contributions calculated
directly from $\Pi ^{\mathrm{OPE}}(p)$. In the present work, we compute $\Pi
(M^{2},s_{0})$ by taking into account nonperturbative terms up to the 10th dimension. 
The explicit expression of the function $\Pi (M^{2},s_{0})$ is given
in Appendix.

The sum rules for the mass $m_{\mathcal{Z}}$ and coupling $f_{\mathcal{Z}}$
of the scalar tetraquark $\mathcal{Z}_{b:\overline{s}}^{0}$ can be found in
the same manner. The correlator $\Pi ^{\mathrm{Phys}}(p)$ contains only a
trivial Lorentz structure proportional to $I$, and the relevant invariant
amplitude has the simple form $\Pi ^{\mathrm{Phys}}(p^{2})=m_{\mathcal{Z}%
}^{2}f_{\mathcal{Z}}^{2}/(m_{\mathcal{Z}}^{2}-p^{2})$. The QCD side of the sum
rules is determined by the formula
\begin{eqnarray}
&&\Pi ^{\mathrm{OPE}}(p)=i\int d^{4}xe^{ipx}\mathrm{Tr}\left[ \gamma _{5}%
\widetilde{S}_{b}^{aa^{\prime }}(x)\gamma _{5}S_{c}^{bb^{\prime }}(x)\right]
\notag \\
&&\times \left\{ \mathrm{Tr}\left[ \gamma _{5}\widetilde{S}_{s}^{b^{\prime
}b}(-x)\gamma _{5}S_{u}^{a^{\prime }a}(-x)\right] -\mathrm{Tr}\left[ \gamma
_{5}\widetilde{S}_{s}^{a^{\prime }b}(-x)\right. \right.  \notag \\
&&\left. \times \gamma _{5}S_{u}^{b^{\prime }a}(-x)\right] -\mathrm{Tr}\left[
\gamma _{5}\widetilde{S}_{s}^{b^{\prime }a}(-x)\gamma _{5}S_{u}^{a^{\prime
}b}(-x)\right]  \notag \\
&&\left. +\mathrm{Tr}\left[ \gamma _{5}\widetilde{S}_{d}^{a^{\prime
}a}(-x)\gamma _{5}S_{u}^{b^{\prime }b}(-x)\right] \right\}.
\label{eq:CFScalar}
\end{eqnarray}%
The parameters of $\mathcal{Z}_{b:\overline{s}}^{0}$ after evident
replacements $\Pi (M^{2},s_{0})\rightarrow \widetilde{\Pi }(M^{2},s_{0})$
and $\mathcal{M}\rightarrow \widetilde{\mathcal{M}}=m_{b}+m_{c}+m_{s}$ are
determined by Eqs.\ (\ref{eq:MassVector}) and (\ref{eq:CouplingV}). Here, $%
\widetilde{\Pi }(M^{2},s_{0})$ is the transformed and subtracted invariant
amplitude corresponding to the correlation function $\Pi ^{\mathrm{OPE}}(p)$.

The sum rules through the propagators depend on different vacuum
condensates. These condensates are universal parameters of computations and
do not depend on the analyzed problem. It is worth noting that the light
quark propagator contains various quark, gluon, and mixed condensates of
different dimensions. Some of these terms, for example, $\langle \overline{q}%
g_{s}\sigma Gq\rangle $ and $\langle \overline{s}g_{s}\sigma Gs\rangle $, $%
\langle \overline{q}q\rangle ^{2}$ and $\langle \overline{s}s\rangle ^{2}$, $%
\langle \overline{q}q\rangle \langle g_{s}G^{2}\rangle $ and $\langle
\overline{s}s\rangle \langle g_{s}G^{2}\rangle $, and others were
obtained from higher-dimensional condensates using the factorization
hypothesis. However, the factorization assumption is not precise and is violated in
the case of higher-dimensional condensates \cite{Ioffe:2005ym}: for the
condensates of dimension 10, even the order of magnitude of such a violation
is unclear. Nevertheless, the contributions of these terms are small; therefore,
in what follows, we ignore the uncertainties generated by this violation. Below,
we list the vacuum condensates and masses of $b$, $c$, and $s$ quarks used
in our numerical analysis:
\begin{eqnarray}
&&\langle \overline{q}q\rangle =-(0.24\pm 0.01)^{3}~\mathrm{GeV}^{3},\
\langle \overline{s}s\rangle =0.8\ \langle \bar{q}q\rangle ,  \notag \\
&&\langle \overline{q}g_{s}\sigma Gq\rangle =m_{0}^{2}\langle \overline{q}%
q\rangle ,\ \langle \overline{s}g_{s}\sigma Gs\rangle =m_{0}^{2}\langle \bar{%
s}s\rangle ,  \notag \\
&&m_{0}^{2}=(0.8\pm 0.1)~\mathrm{GeV}^{2}  \notag \\
&&\langle \frac{\alpha _{s}G^{2}}{\pi }\rangle =(0.012\pm 0.004)~\mathrm{GeV}%
^{4},  \notag \\
&&\langle g_{s}^{3}G^{3}\rangle =(0.57\pm 0.29)~\mathrm{GeV}^{6},\
m_{s}=93_{-5}^{+11}~\mathrm{MeV},  \notag \\
&&m_{c}=1.27\pm 0.2~\mathrm{GeV},\ m_{b}=4.18_{-0.02}^{+0.03}~\mathrm{GeV}.
\label{eq:Parameters}
\end{eqnarray}%
In Eq.\ (\ref{eq:Parameters}), we introduced the following short-hand notations:
\begin{equation}
G^{2}=G_{\alpha \beta }^{A}G_{\alpha \beta }^{A},\ G^{3}=f^{ABC}G_{\alpha
\beta }^{A}G_{\beta \delta }^{B}G_{\delta \alpha }^{C},
\end{equation}%
where $G_{\alpha \beta }^{A}$ is the gluon field strength tensor, $f^{ABC}$
are the structure constants of the color group $SU_{c}(3)$, and $%
A,B,C=1,2,...8$.

The mass and coupling of the tetraquarks (\ref{eq:MassVector}) and (\ref%
{eq:CouplingV}) also depend on the Borel and continuum threshold parameters $%
M^{2}$ and $s_{0}$. The $M^{2}$ and $s_{0}$ are the auxiliary quantities,
and their correct choice is one of the important problems in sum rule studies.
Proper working regions for $M^{2}$ and $s_{0}$ must satisfy restrictions
imposed on the pole contribution ($\mathrm{PC}$) and convergence of the
operator product expansion measured by the ratio $R(M^{2})$, which we define
respectively by the expressions
\begin{equation}
\mathrm{PC}=\frac{\Pi (M^{2},s_{0})}{\Pi (M^{2},\infty )},  \label{eq:PC}
\end{equation}%
and
\begin{equation}
R(M^{2})=\frac{\Pi ^{\mathrm{DimN}}(M^{2},s_{0})}{\Pi (M^{2},s_{0})}.
\label{eq:Convergence}
\end{equation}%
Here, $\Pi ^{\mathrm{DimN}}(M^{2},s_{0})$ is a contribution to the
correlation function of the last term (or sum of the last few terms) in the
operator product expansion. In the present work, we use the following
restrictions imposed on these parameters: at the maximal edge of $M^{2}$, the
pole contribution should obey $\mathrm{PC}>0.2$, and at the minimum of $%
M^{2}$, we require fulfilment of $R(M^{2})\leq 0.01$. Lets us note that we
estimate $R(M^{2})$ using the last three terms in the OPE $\mathrm{DimN}=\mathrm{%
Dim(8+9+10)}$.

Variations of $M^{2}$ and $s_{0}$ within the allowed working regions are
the main sources of theoretical errors in sum rule computations. Therefore, the
Borel parameter $M^{2}$ should be fixed for minimizing the
dependence of extracted physical quantities on its variations. The situation
with $s_{0}$ is more subtle, because it bears physical information about the
excited states of the tetraquark $T_{b:\overline{s}}^{\mathrm{AV}}$. In
fact, the continuum threshold parameter $s_{0}$ separates the ground-state
contribution from the ones of higher resonances and continuum states; hence, $%
s_{0}$ should be below the first excitation of $T_{b:\overline{s}}^{\mathrm{AV}%
}$. However, available information on the excited states of tetraquarks is limited to
only a few theoretical studies \cite{Maiani:2014,Wang:2014vha,Agaev:2017tzv}. As a
result, one fixes $s_{0}$ to achieve maximal for $\mathrm{PC}$, ensuring
fulfilment of the other constraints and simultaneously keeping the computation
self-consistency under control. The latter means that the gap $\sqrt{%
s_{0}}-m_{\mathrm{AV}}$ in the case of heavy tetraquarks should be  $%
\sim 600~\mathrm{MeV}$, which serves as a measure of excitation.

Numerical analysis suggests that regions
\begin{equation}
M^{2}\in \lbrack 9,12]\ \mathrm{GeV}^{2},\ s_{0}\in \lbrack 115,120]\
\mathrm{GeV}^{2},  \label{eq:Wind1}
\end{equation}%
satisfy all of the aforementioned constraints on $M^{2}$ and $s_{0}$. Thus, at $%
M^{2}=12~\mathrm{GeV}^{2}$, the pole contribution is $0.23$, and at $M^{2}=9~%
\mathrm{GeV}^{2}$, it amounts to $0.62$. These values of $M^{2}$ limit the
boundaries of a region in which the Borel parameter can be changed. At the
minimum of $M^{2}=9~\mathrm{GeV}^{2}$, we get $R\approx 0.005$. In addition,
at the minimum of the Borel parameter, the perturbative contribution is $79\%$
of the result overshooting the nonperturbative effects.

For $m_{\mathrm{AV}}$ and $f_{\mathrm{AV}}$, we have obtained
\begin{eqnarray}
m_{\mathrm{AV}} &=&(10215\pm 250)~\mathrm{MeV},  \notag \\
f_{\mathrm{AV}} &=&(2.26\pm 0.57)\times 10^{-2}~\mathrm{GeV}^{4}.
\label{eq:Result1}
\end{eqnarray}%
In Eq.\ (\ref{eq:Result1}), the theoretical uncertainties of computations are
shown as well. For the mass $m_{\mathrm{AV}}$, these uncertainties are $%
\pm 2.4\%$ of the central value, and for the coupling $f_{\mathrm{AV}}$,
they amount to $\pm 25\%$, but in both cases, they remain within the limits accepted
by the sum rule computations. In Fig.\ \ref{fig:MassVector}, we plot our
prediction for $m_{\mathrm{AV}}$ as a function of $M^{2}$ and $s_{0}$: one
can see a mild dependence of $m_{\mathrm{AV}}$ on these parameters. It is
also evident that
\begin{equation}
\sqrt{s_{0}}-m_{\mathrm{AV}}=\left[ 510,740\right] ~\mathrm{MeV,}
\label{eq:Massgap}
\end{equation}%
which is a reasonable mass gap between the ground-state and excited heavy
tetraquarks.

Returning to the issue of the two-meson continuum, we can now compare the mass
of the tetraquark $T_{b:\overline{s}}^{\mathrm{AV}}$ with the energy level of
this continuum. It is clear that the two-meson continuum may be populated
by pairs $B^{-}B_{s}^{\ast }$ and $B^{\ast -}\overline{B}_{s}^{0}$, and that
$T_{b:\overline{s}}^{\mathrm{AV}}$ is $\approx 480~\mathrm{MeV}$ below it.
This difference is comparable to (\ref{eq:Massgap}); hence, one can ignore
the two-meson continuum's impact on the physical parameters of $T_{b:\overline{s}%
}^{\mathrm{AV}}$.

The mass $m_{\mathcal{Z}}$ and coupling $f_{\mathcal{Z}}$ of the state $%
\mathcal{Z}_{b:\overline{s}}^{0}$ are found from the sum rules by utilizing
the following working windows for $M^{2}$ and $s_{0}$ 
\begin{equation}
M^{2}\in \lbrack 5.5,6.5]~\mathrm{GeV}^{2},\ s_{0}\in \lbrack 52,54]~\mathrm{%
GeV}^{2}.  \label{eq:Wind2}
\end{equation}%
The regions (\ref{eq:Wind2}) satisfy standard restrictions associated with the sum rule
computations. In fact, at $M^{2}=5.5~\mathrm{GeV}^{2}$, the ratio $R$ is $%
0.009$; hence, the convergence of the sum rules is satisfied. The pole
contribution $\mathrm{PC}$ at $M^{2}=6.5~\mathrm{GeV}^{2}$ and $M^{2}=5.5~%
\mathrm{GeV}^{2}$ equals to $0.23$ and $0.61$, respectively. At the minimum of $%
M^{2}$, the perturbative contribution constitutes $72\%$ of the entire result
and considerably exceeds that of nonperturbative terms.

For $m_{\mathcal{Z}}$ and $f_{\mathcal{Z}}$, our computations yield
\begin{eqnarray}
m_{\mathcal{Z}} &=&(6770\pm 150)~\mathrm{MeV},  \notag \\
f_{\mathcal{Z}} &=&(6,3\pm 1.3)\times 10^{-3}~\mathrm{GeV}^{4}.
\label{eq:Result2}
\end{eqnarray}%
In Fig.\ \ref{fig:MassZ}, we depict the mass of the tetraquark $\mathcal{Z}%
_{b:\overline{s}}^{0}$ and demonstrate its dependence on $M^{2}$ and $s_{0}$.

The mass of the axial-vector tetraquark $T_{b:\overline{s}}^{\mathrm{AV}}$
was calculated in Ref.\ \cite{Du:2012wp}\ in the context of the QCD sum rule
method, using different interpolating currents. Computations were
performed with dimension $8$ accuracy, and two lowest predictions for the mass
of the axial-vector particle $bb\overline{q}\overline{s}$ were obtained within
ranges $(10300\pm 300)~\mathrm{MeV}$ and $(10300\pm 400)~\mathrm{MeV}$. Our
result is close to the central value of these predictions. The difference in
theoretical errors can be attributed to the higher accuracy of our
computations and more detailed quark propagators used in analysis. The
authors of Ref.\ \cite{Du:2012wp} noted the strong interaction stable nature of $%
T_{b:\overline{s}}^{\mathrm{AV}}$. As we will see below, our investigation
proves that $T_{b:\overline{s}}^{\mathrm{AV}}$ is stable against strong and
radiative decays and can transform only through weak processes.

The scalar tetraquark with the quark content $[bc][\overline{u}\overline{s}]$
was explored recently in Ref.\ \cite{Wang:2020jgb}. The predicted mass
of this state $(7.14\pm 0.12)~\mathrm{GeV}$ obtained there is larger than
our prediction (\ref{eq:Result2}). Such a sizeable difference between the two
results can be explained by some factors. Thus, in the present work,
calculations have been performed by taking into account dimension $\ 10$
condensates, whereas in Ref.\ \cite{Wang:2020jgb}, the authors included 
nonperturbative terms up the to eighth dimension into analysis. We have used more
detailed expressions for quark propagators, including the terms $\sim
g_{s}^{2}\langle \overline{q}q\rangle ^{2}$ and $\sim \langle \overline{q}%
q\rangle \langle g_{s}^{2}G^{2}\rangle $ in the light and $\sim \langle
g_{s}^{3}G^{3}\rangle $ in the heavy quark propagators. However, in our view,
the choice of the working windows for the parameters $M^{2}$ and $s_{0}$ is the main
source of fixed discrepancies. The regions for $M^{2}$ and $s_{0}$ should
be extracted from the analysis of constraints (\ref{eq:PC}) and (\ref%
{eq:Convergence}) imposed on the invariant amplitude $\Pi (M^{2},s_{0})$.
The $\mathrm{PC}$ in the present investigation varies within limits $%
0.61-0.23$, which corresponds to the boundaries of the Borel region. Let us
emphasize that we extract the parameters $m_{\mathcal{Z}}$ and $f_{\mathcal{Z}}$
approximately in the middle region of the window (\ref{eq:Wind2}), where the
pole contribution is $\mathrm{PC}\approx 0.42-0.45$. The working regions for
$M^{2}$ and $s_{0}$ used in Ref.\ \cite{Wang:2020jgb} ensure only $\mathrm{PC%
}\approx 0.31$, which may generate differences in the extracted values of $m_{%
\mathcal{Z}}$.


\section{Weak form factors $G_{i}(p^{2})$ and semileptonic decays $T_{b:%
\overline{s}}^{\mathrm{AV}}\rightarrow \mathcal{Z}_{b:\overline{s}}^{0}l%
\overline{\protect\nu }_{l}$}

\label{sec:Decays1}
The analysis performed in the previous section confirms that the tetraquark $%
T_{b:\overline{s}}^{\mathrm{AV}}$ is stable against the strong and
electromagnetic decays. Indeed, the mass of this state $m_{\mathrm{AV}%
}=10215~\mathrm{MeV}$ is $480/477~\mathrm{MeV}$ below the thresholds $%
10695/10692~\mathrm{MeV}$ for its strong decays to mesons $B^{-}B_{s}^{\ast
} $ and $B^{\ast -}\overline{B}_{s}^{0}$, respectively. The maximum of the
mass $10465~\mathrm{MeV}$ is still below these limits. The threshold $10646~%
\mathrm{MeV}$ for the process $T_{b:\overline{s}}^{\mathrm{AV}}\rightarrow
B^{-}\overline{B}_{s}^{0}\gamma $ also exceeds the maximal allowed value of $%
m_{\mathrm{AV}}$, which forbids this electromagnetic decay. Therefore, the
full width and mean lifetime of $T_{b:\overline{s}}^{\mathrm{AV}}$ are
determined by its weak decays.

\begin{figure}[h!]
\begin{center}
\includegraphics[totalheight=6cm,width=8cm]{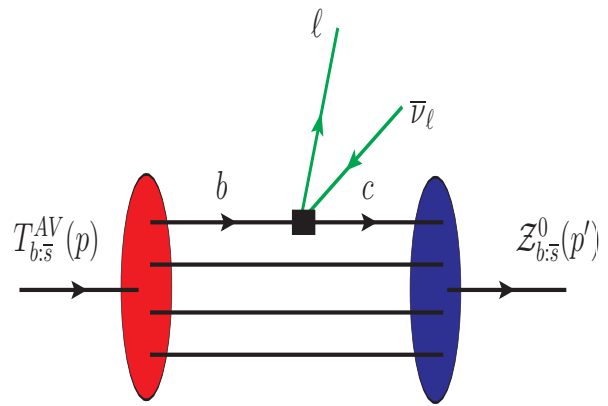}
\end{center}
\caption{The Feynman diagram for the semileptonic decay $T_{b:\overline{s}}^{%
\mathrm{AV}}\rightarrow \mathcal{Z}_{b:\overline{s}}^{0}l\overline{\protect%
\nu }_{l}$. The black square denotes the effective weak vertex.}
\label{fig:Decay1}
\end{figure}

There are different weak decay channels of $T_{b:\overline{s}}^{\mathrm{AV}}$%
, which can be generated by sub-processes $b\rightarrow W^{-}c$ and $%
b\rightarrow W^{-}u$. The decays triggered by the transition $b\rightarrow
W^{-}c$ are dominant processes relative to the ones connected with $b\rightarrow
W^{-}u$: the latter decays are suppressed relative to the dominant decays by a factor $%
|V_{bu}|^{2}/|V_{bc}|^{2}$ $\simeq 0.01$, with $V_{q_{1}q_{2}}$ being the
Cabibbo-Khobayasi-Maskawa (CKM) matrix elements. In the present work, we
restrict ourselves to the analysis of the dominant weak decays of $T_{b:%
\overline{s}}^{\mathrm{AV}}$ (see Fig.\ \ref{fig:Decay1}).

The dominant processes themselves can be categorized into two groups: the
first group contains the semileptonic decays $T_{b:\overline{s}}^{\mathrm{AV}%
}\rightarrow \mathcal{Z}_{b:\overline{s}}^{0}l\overline{\nu }_{l}$ , whereas
the nonleptonic transitions $T_{b:\overline{s}}^{\mathrm{AV}}\rightarrow
\mathcal{Z}_{b:\overline{s}}^{0}M$ belong to the second group. In this
section, we consider the semileptonic decays and calculate the partial widths
of the processes $T_{b:\overline{s}}^{\mathrm{AV}}\rightarrow \mathcal{Z}_{b:%
\overline{s}}^{0}l\overline{\nu }_{l}$, where $l$ is one of the lepton
species $e, \mu $ and $\tau $. Owing to the large mass difference between the
initial and final tetraquarks, $3445~\mathrm{MeV}$, all of these semileptonic
decays are kinematically allowed ones.

The effective Hamiltonian to describe the subprocess $b\rightarrow W^{-}c$ \
at the tree-level is given by the expression
\begin{equation}
\mathcal{H}^{\mathrm{eff}}=\frac{G_{F}}{\sqrt{2}}V_{bc}\overline{c}\gamma
_{\mu }(1-\gamma _{5})b\overline{l}\gamma ^{\mu }(1-\gamma _{5})\nu _{l},
\label{eq:EffecH}
\end{equation}%
with $G_{F}$ and $V_{bc}$ being the Fermi coupling constant and CKM matrix
element, respectively. A matrix element of $\mathcal{H}^{\mathrm{eff}}$
between the initial and final tetraquarks is equal to
\begin{equation}
\langle \mathcal{Z}_{b:\overline{s}}^{0}(p^{\prime })|\mathcal{H}^{\mathrm{%
eff}}|T_{b:\overline{s}}^{\mathrm{AV}}(p)\rangle =L_{\mu }H^{\mu },
\end{equation}%
where $L_{\mu }$ and $H^{\mu }$ are the leptonic and hadronic factors,
respectively. A treatment of $L_{\mu }$ is trivial; therefore, we consider
the matrix element $H^{\mu }$ in a detailed form, which depends on the parameters
of the tetraquarks. After factoring out the constant factors, $H^{\mu }$ is
the matrix element of the current
\begin{equation}
J_{\mu }^{\mathrm{tr}}=\overline{c}\gamma _{\mu }(1-\gamma _{5})b.
\label{eq:TrCurr}
\end{equation}%
The matrix element $\langle \mathcal{Z}_{b:\overline{s}}^{0}(p^{\prime
})|J_{\mu }^{\mathrm{tr}}|T_{b:\overline{s}}^{\mathrm{AV}}(p)\rangle $
describes the weak transition of the axial-vector tetraquark to the scalar
particle and is expressible in terms of four weak form factors $%
G_{i}(q^{2}) $ that parametrize long-distance dynamical effects of this
transformation \cite{Wirbel:1985ji,Ball:1991bs}
\begin{eqnarray}
&&\langle \mathcal{Z}_{b:\overline{s}}^{0}(p^{\prime })|J_{\mu }^{\mathrm{tr}%
}|T_{b:\overline{s}}^{\mathrm{AV}}(p)\rangle =\widetilde{G}%
_{0}(q^{2})\epsilon _{\mu }+\widetilde{G}_{1}(q^{2})(\epsilon p^{\prime
})P_{\mu }  \notag \\
&&+\widetilde{G}_{2}(q^{2})(\epsilon p^{\prime })q_{\mu }+i\widetilde{G}%
_{3}(q^{2})\varepsilon _{\mu \nu \alpha \beta }\epsilon ^{\nu }p^{\alpha
}p^{\prime }{}^{\beta }.  \label{eq:Vertex}
\end{eqnarray}%
The scaled functions $\widetilde{G}_{i}(q^{2})$ are connected with the
dimensionless form factors $G_{i}(q^{2})$ by the equalities
\begin{equation}
\widetilde{G}_{0}(q^{2})=\widetilde{m}G_{0}(q^{2}),\ \widetilde{G}%
_{j}(q^{2})=\frac{G_{j}(q^{2})}{\widetilde{m}},\ j=1,2,3.  \label{eq:VertexA}
\end{equation}%
Here, $\widetilde{m}=m_{\mathrm{AV}}+m_{\mathcal{Z}}$, $p_{\mu }$ and $%
\epsilon _{\mu }$ are the momentum and polarization vector of the tetraquark
$T_{b:\overline{s}}^{\mathrm{AV}}$, $p^{\prime }$ is the momentum of the
scalar state $\mathcal{Z}_{b:\overline{s}}^{0}$. We use also $P_{\mu
}=p_{\mu }^{\prime }+p_{\mu }$ and $q_{\mu }=p_{\mu }-p_{\mu }^{\prime }$,
the latter being the momentum transferred to the leptons. It is evident that $%
q^{2}$ varies within  $m_{l}^{2}\leq q^{2}\leq (m_{\mathrm{AV}%
}-m_{\mathcal{Z}})^{2},$ where $m_{l}$ is the mass of a lepton $l$.

The sum rules for the form factors $G_{i}(q^{2})$ can be obtained by
analyzing the three-point correlation function
\begin{eqnarray}
\Pi _{\mu \nu }(p,p^{\prime }) &=&i^{2}\int d^{4}xd^{4}ye^{i(p^{\prime
}y-px)}  \notag \\
&&\times \langle 0|\mathcal{T}\{J_{\mathcal{Z}}(y)J_{\nu }^{\mathrm{tr}%
}(0)J_{\mu }^{^{\dagger }}(x)\}|0\rangle .  \label{eq:CF7}
\end{eqnarray}%
To this end, we have to express $\Pi _{\mu \nu }(p,p^{\prime })$ using the
masses and couplings of the tetraquarks and thus determine the
physical side of the sum rules $\Pi _{\mu \nu }^{\mathrm{Phys}}(p,p^{\prime
})$. The function $\Pi _{\mu \nu }^{\mathrm{Phys}}(p,p^{\prime })$ can be
presented as%
\begin{eqnarray}
&&\Pi _{\mu \nu }^{\mathrm{Phys}}(p,p^{\prime })=\frac{\langle 0|J_{\mathcal{%
Z}}|\mathcal{Z}_{b:\overline{s}}^{0}(p^{\prime })\rangle \langle \mathcal{Z}%
_{b:\overline{s}}^{0}(p^{\prime })|J_{\nu }^{\mathrm{tr}}|T_{b:\overline{s}%
}^{\mathrm{AV}}(p,\epsilon )\rangle }{(p^{2}-m_{\mathrm{AV}}^{2})(p^{\prime
2}-m_{\mathcal{Z}}^{2})}  \notag \\
&&\times \langle T_{b:\overline{s}}^{\mathrm{AV}}(p,\epsilon )|J_{\mu
}^{^{\dagger }}|0\rangle +\cdots,  \label{eq:CF7a}
\end{eqnarray}%
where we take into account the contribution of the ground-state
particles and denote the effects of the excited and continuum states by dots.

The phenomenological side of the sum rules can be simplified by substituting
into Eq.\ (\ref{eq:CF7a}) the expressions of matrix elements in terms of the
tetraquarks' masses and couplings as well as weak transition form factors. For
these purposes, we employ Eqs.\ (\ref{eq:MElem1}) and (\ref{eq:Vertex}) and
define the matrix element of $\mathcal{Z}_{b:\overline{s}}^{0}$
\begin{equation}
\langle 0|J_{\mathcal{Z}}|\mathcal{Z}_{b:\overline{s}}^{0}(p^{\prime
})\rangle =f_{\mathcal{Z}}m_{\mathcal{Z}}.  \label{eq:ME3}
\end{equation}%
Then, one gets
\begin{eqnarray}
&&\Pi _{\mu \nu }^{\mathrm{Phys}}(p,p^{\prime })=\frac{f_{\mathrm{AV}}m_{%
\mathrm{AV}}f_{\mathcal{Z}}m_{\mathcal{Z}}}{(p^{2}-m_{\mathrm{AV}%
}^{2})(p^{\prime 2}-m_{\mathcal{Z}}^{2})}  \notag \\
&&\times \left\{ \widetilde{G}_{0}(q^{2})\left( -g_{\mu \nu }+\frac{p_{\mu
}p_{\nu }}{m_{\mathrm{AV}}^{2}}\right) +\left[ \widetilde{G}%
_{1}(q^{2})P_{\mu }\right. \right.  \notag \\
&&\left. +\widetilde{G}_{2}(q^{2})q_{\mu }\right] \left( -p_{\nu }^{\prime }+%
\frac{m_{\mathrm{AV}}^{2}+m_{\mathcal{Z}}^{2}-q^{2}}{2m_{\mathrm{AV}}^{2}}%
p_{\nu }\right)  \notag \\
&&\left. -i\widetilde{G}_{3}(q^{2})\varepsilon _{\mu \nu \alpha \beta
}p^{\alpha }p^{\prime }{}^{\beta }\right\} +\cdots.  \label{eq:Phys2}
\end{eqnarray}

We should also calculate the correlation function in terms of the quark
propagators and find $\Pi _{\mu \nu }^{\mathrm{OPE}}(p,p^{\prime })$. \ The
function $\Pi _{\mu \nu }^{\mathrm{OPE}}(p,p^{\prime })$ is the second side
of the sum rules and has the following form
\begin{eqnarray}
&&\Pi _{\mu \nu }^{\mathrm{OPE}}(p,p^{\prime })=\int
d^{4}xd^{4}ye^{i(p^{\prime }y-px)}\left\{ \mathrm{Tr}\left[ \gamma _{5}%
\widetilde{S}_{s}^{ba^{\prime }}(x-y)\right. \right.  \notag \\
&&\left. \times \gamma _{5}S_{u}^{a^{\prime }b}(x-y)\right] \left( \mathrm{Tr%
}\left[ \gamma _{\mu }\widetilde{S}_{b}^{aa^{\prime }}(y-x)\gamma
_{5}S_{c}^{bi}(y)\gamma _{\nu }(1-\gamma _{5})\right. \right.  \notag \\
&&\left. \times S_{b}^{ib^{\prime }}(-x)\right] +\mathrm{Tr}\left[ \gamma
_{\mu }\widetilde{S}_{b}^{ia^{\prime }}(-x)(1-\gamma _{5})\gamma _{\nu }%
\widetilde{S}_{c}^{bi}(y)\gamma _{5}\right.  \notag \\
&&\left. \left. \times S_{b}^{ab^{\prime }}(y-x)\right] \right) -\mathrm{Tr}%
\left[ \gamma _{5}\widetilde{S}_{s}^{b^{\prime }a}(x-y)\gamma
_{5}S_{u}^{a^{\prime }b}(x-y)\right]  \notag \\
&&\times \left( \mathrm{Tr}\left[ \gamma _{\mu }\widetilde{S}%
_{b}^{aa^{\prime }}(y-x)\gamma _{5}S_{c}^{bi}(y)\gamma _{\nu }(1-\gamma
_{5})S_{b}^{ib^{\prime }}(-x)\right] \right.  \notag \\
&&\left. \left. +\mathrm{Tr}\left[ \gamma _{\mu }\widetilde{S}%
_{b}^{ia^{\prime }}(-x)(1-\gamma _{5})\gamma _{\nu }\widetilde{S}%
_{c}^{bi}(y)\gamma _{5}S_{b}^{ab^{\prime }}(y-x)\right] \right) \right\}.
\notag \\
&&  \label{eq:DecayCF}
\end{eqnarray}

To extract expressions of the form factors $G_{i}(q^{2})$, we
equate invariant amplitudes corresponding to the same Lorentz structures
both in $\Pi _{\mu \nu }^{\mathrm{Phys}}(p,p^{\prime })$ and $\Pi _{\mu \nu
}^{\mathrm{OPE}}(p,p^{\prime })$, carry out double Borel transformations
over the variables $p^{\prime 2}$ and $p^{2}$, and perform continuum
subtraction. For instance, to extract the sum rule for $\widetilde{G}%
_{0}(q^{2})$, we use the structure $g_{\mu \nu }$, whereas for $\widetilde{G}%
_{3}(q^{2})$, the term $\sim \varepsilon _{\mu \nu \alpha \beta }p^{\alpha
}p^{\prime }{}^{\beta }$ can be employed. The sum rules for the scaled form
factors $\widetilde{G}_{i}(q^{2})$ can be written in a single formula
\begin{eqnarray}
&&\widetilde{G}_{i}(\mathbf{M}^{2},\mathbf{s}_{0},q^{2})=\frac{1}{f_{\mathrm{%
AV}}m_{\mathrm{AV}}f_{\mathcal{Z}}m_{\mathcal{Z}}}\int_{\mathcal{M}%
^{2}}^{s_{0}}dse^{(m_{\mathrm{AV}}^{2}-s)/M_{1}^{2}}  \notag \\
&&\times \int_{\widetilde{\mathcal{M}}^{2}}^{s_{0}^{\prime }}ds^{\prime
}\rho _{i}(s,s^{\prime })e^{(m_{\mathcal{Z}}^{2}-s^{\prime })/M_{2}^{2}},
\label{eq:FF}
\end{eqnarray}%
where $\rho _{i}(s,s^{\prime })$ are spectral densities computed as the
imaginary parts of the corresponding terms in $\Pi _{\mu \nu }^{\mathrm{OPE}%
}(p,p^{\prime })$. They contain perturbative and nonperturbative
contributions and are found in the present work with dimension-6 accuracy.
In Eq.\ (\ref{eq:FF}), $\mathbf{M}^{2}=(M_{1}^{2},M_{2}^{2})$ and $\mathbf{s}%
_{0}=(s_{0},\ s_{0}^{\prime })$ are the Borel and continuum threshold
parameters, respectively. The pair of parameters ($M_{1}^{2}$, $s_{0}$)
corresponds to the initial tetraquark's channels, whereas ($M_{2}^{2}$, $%
s_{0}^{\prime }$) describes the final-state tetraquark.

As usual, the form factors $\widetilde{G}_{i}(\mathbf{M}^{2},\mathbf{s}%
_{0},q^{2})$ contain various input parameters, which should be determined
before numerical analysis. The vacuum condensates of quark, gluon, and mixed
operators are already presented in Eq.\ (\ref{eq:Parameters}). The masses
and couplings of the tetraquarks $T_{b:\overline{s}}^{\mathrm{AV}}$ and $%
\mathcal{Z}_{b:\overline{s}}^{0}$ have been extracted in Section \ref%
{sec:Masses}. The Borel and continuum threshold parameters $\mathbf{M}^{2}$
and $\mathbf{s}_{0}$ should be chosen so as to meet all
restrictions of sum rule computations. One has also to bear in mind that $%
\widetilde{G}_{i}(\mathbf{M}^{2},\mathbf{s}_{0},q^{2})$ depends on the masses and
couplings of the initial and final tetraquarks, which have been evaluated
also in the context of the sum rule approach. We fix the auxiliary
parameters ($M_{1}^{2}$, $s_{0}$) and ($M_{2}^{2}$, $s_{0}^{\prime }$) as in
the corresponding mass computations, because they satisfy standard constraints
of three-point sum rule calculations and do not generate additional
uncertainties in the spectroscopic parameters of relevant tetraquarks.

\begin{figure}[h]
\includegraphics[width=8.8cm]{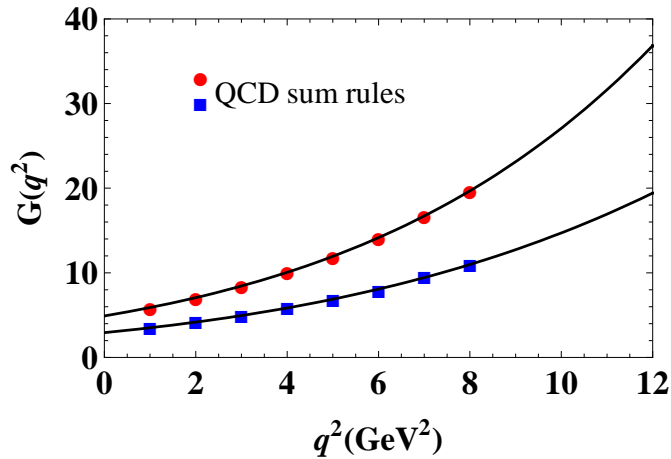}
\caption{  Sum rule results for the form factors $G_{0}(q^{2})$ (red
circles) and $G_{1}(q^{2})$ (blue squares). The solid curves  are fit functions $\mathcal{G}_{0}(q^{2})$ and $\mathcal{G}_{1}(q^{2})$
.}
\label{fig:AVWFF}
\end{figure}

The form factors $\widetilde{G}_{i}(q^{2})$ determine the differential decay
rate $d\Gamma /dq^{2}$ of the semileptonic decay $T_{b:\overline{s}}^{%
\mathrm{AV}}\rightarrow \mathcal{Z}_{b:\overline{s}}^{0}l\overline{\nu }_{l}$%
, the explicit expression of which can be found in Ref. \cite{Agaev:2018khe}%
. The partial width of the process is equal to an integral of this rate over
the momentum transfer $q^{2}$ within the limits $m_{l}^{2}\leq q^{2}\leq (m_{%
\mathrm{AV}}-m_{\mathcal{Z}})^{2}$. Our results for the form factors are
plotted in Fig.\ \ref{fig:AVWFF}. The QCD sum rules lead to reliable
predictions at $m_{l}^{2}\leq q^{2}\leq 8~\mathrm{GeV}^{2}$. However, these predictions
do not cover the entire integration region $m_{l}^{2}\leq q^{2}\leq 11.87~%
\mathrm{GeV}^{2}$. To solve this problem, one has to introduce extrapolation
functions $\mathcal{G}_{i}(q^{2})$ of a relatively simple analytic form, which,
at $q^{2}$ accessible to the QCD sum rules, coincide with their predictions but
can be used in the entire region.

For these purposes, we choose to work with the functions
\begin{equation}
\mathcal{G}_{i}(q^{2})=\mathcal{G}_{0}^{i}\exp \left[ g_{1}^{i}\frac{q^{2}}{%
m_{\mathrm{AV}}^{2}}+g_{2}^{i}\left( \frac{q^{2}}{m_{\mathrm{AV}}^{2}}%
\right) ^{2}\right],  \label{eq:FFunctions}
\end{equation}%
in which the parameters $\mathcal{G}_{0}^{i},~g_{1}^{i}$, and $g_{2}^{i}$ should
be fitted to satisfy the sum rules' predictions. The parameters of the functions
$\mathcal{G}_{i}(q^{2})$, obtained numerically, are presented in
Table\ \ref{tab:ExtraF}.
\begin{table}[tbp]
\begin{tabular}{|c|c|c|c|}
\hline
$\mathcal{G}_{i}(q^{2})$ & $\mathcal{G}_{0}^{i}$ & $g_{1}^{i}$ & $g_{2}^{i}$
\\ \hline
$\mathcal{G}_{0}(q^{2})$ & $4.91$ & $19.29$ & $-15.34$ \\
$\mathcal{G}_{1}(q^{2})$ & $2.94$ & $18.73$ & $-20.09$ \\
$\mathcal{G}_{2}(q^{2})$ & $-22.67$ & $20.50$ & $-22.95$ \\
$\mathcal{G}_{3}(q^{2})$ & $-21.14$ & $20.77$ & $-23.62$ \\ \hline\hline
\end{tabular}%
\caption{Parameters of the extrapolating functions $\mathcal{G}_{i}(q^{2})$.}
\label{tab:ExtraF}
\end{table}
The functions $\mathcal{G}_{i}(q^{2})$ are also shown in Fig. \ref{fig:AVWFF}
: one can see a good agreement between the sum rule predictions and
fit functions.

In the numerical computations for the Fermi constant, CKM matrix elements, and
masses of leptons, we use
\begin{eqnarray}
G_{F} &=&1.16637\times 10^{-5}~\mathrm{GeV}^{-2},  \notag \\
|V_{bc}| &=&(42.2\pm 0.08)\times 10^{-3}.
\end{eqnarray}%
$m_{e}=0.511~\mathrm{MeV}$, $m_{\mu }=105.658~\mathrm{MeV}$, and $%
m_{\tau }=(1776.82~\pm 0.16)~\mathrm{MeV}$ \cite{Tanabashi:2018oca}$\mathrm{.%
}$ The predictions obtained for the partial widths of the semileptonic decays
$T_{b:\overline{s}}^{\mathrm{AV}}\rightarrow \mathcal{Z}_{b:\overline{s}%
}^{0}l\overline{\nu }_{l}$ are written as%
\begin{eqnarray}
\Gamma (T_{b:\overline{s}}^{\mathrm{AV}} &\rightarrow &\mathcal{Z}_{b:%
\overline{s}}^{0}e^{-}\overline{\nu }_{e})=(5.34\pm 1.43)\times 10^{-8}~%
\mathrm{MeV},  \notag \\
\Gamma (T_{b:\overline{s}}^{\mathrm{AV}} &\rightarrow &\mathcal{Z}_{b:%
\overline{s}}^{0}\mu ^{-}\overline{\nu }_{\mu })=(5.32\pm 1.41)\times
10^{-8}~\mathrm{MeV},  \notag \\
\Gamma (T_{b:\overline{s}}^{\mathrm{AV}} &\rightarrow &\mathcal{Z}_{b:%
\overline{s}}^{0}\tau ^{-}\overline{\nu }_{\tau })=(2.15\pm 0.54)\times
10^{-8}~\mathrm{MeV},  \notag \\
&&  \label{eq:Results}
\end{eqnarray}%
and are main results of this section.


\section{ Nonleptonic decays $T_{b:\overline{s}}^{\mathrm{AV}}\rightarrow%
\mathcal{Z}_{b:\overline{s}}^{0}M$}

\label{sec:Decays2}

The second class of the weak decays of the tetraquark $T_{b:\overline{s}}^{%
\mathrm{AV}}$ are the processes $T_{b:\overline{s}}^{\mathrm{AV}}\rightarrow
\mathcal{Z}_{b:\overline{s}}^{0}M$, which may affect the full width and
lifetime of the tetraquark $T_{b:\overline{s}}^{\mathrm{AV}}$. Here, we
study the nonleptonic weak decays $T_{b:\overline{s}}^{\mathrm{AV}%
}\rightarrow \mathcal{Z}_{b:\overline{s}}^{0}M$ of the tetraquark $T_{b:%
\overline{s}}^{\mathrm{AV}}$ in the framework of the QCD factorization
method. This approach was applied for investigating the nonleptonic decays of 
conventional mesons \cite{Beneke:1999br,Beneke:2000ry}  but can be also used for
investigating the decays of tetraquarks. Thus, the nonleptonic decays of
scalar exotic mesons $T_{b:\overline{s}}^{-}$, $T_{b:\overline{s}}^{-}$,
$Z_{bc}^{0}$, and $T_{bs;\overline{u}\overline{d}}^{-}$ were explored using
this approach in Refs.\ \cite%
{Agaev:2019lwh,Agaev:2020dba,Sundu:2019feu,Agaev:2019wkk}, respectively.
The weak decays of double- and fully-heavy tetraquarks were analyzed in Refs.\
\cite{Li:2018bkh,Li:2019uch} .

\begin{figure}[h!]
\begin{center}
\includegraphics[totalheight=6cm,width=8cm]{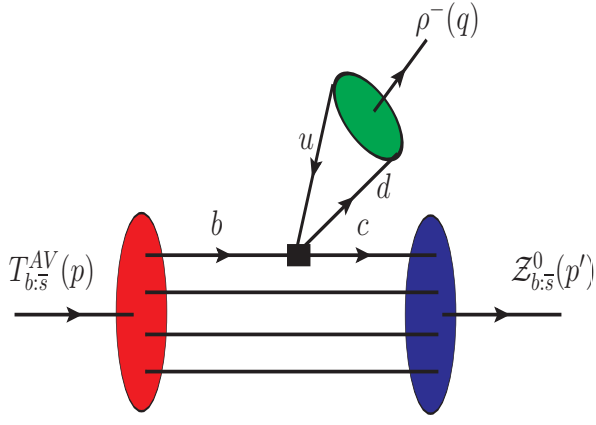}
\end{center}
\caption{ The diagram for the nonleptonic decay $T_{b:\overline{s}}^{\mathrm{%
AV}}\rightarrow \mathcal{Z}_{b:\overline{s}}^{0}\protect\rho ^{-}$.}
\label{fig:Decay2}
\end{figure}

We consider processes where $M$ is one of the vector mesons $\rho ^{-}$, $%
K^{\ast }(892)$, $\ D^{\ast }(2010)^{-}$, and $\ D_{s}^{\ast -}$. We provide
details of analysis for the decay $T_{b:\overline{s}}^{\mathrm{AV}%
}\rightarrow \mathcal{Z}_{b:\overline{s}}^{0}\rho ^{-}$ and write the
final predictions for other channels. The relevant Feynman diagram is shown
in Fig.\ \ref{fig:Decay2}.

At the quark level, the effective Hamiltonian for this decay is given by the
expression
\begin{equation}
\mathcal{H}_{\mathrm{n.-lep}}^{\mathrm{eff}}=\frac{G_{F}}{\sqrt{2}}%
V_{bc}V_{ud}^{\ast }\left[ c_{1}(\mu )Q_{1}+c_{2}(\mu )Q_{2}\right] ,
\label{eq:EffHam}
\end{equation}%
where%
\begin{eqnarray}
Q_{1} &=&\left( \overline{d}_{i}u_{i}\right) _{\mathrm{V-A}}\left( \overline{%
c}_{j}b_{j}\right) _{\mathrm{V-A}},  \notag \\
Q_{2} &=&\left( \overline{d}_{i}u_{j}\right) _{\mathrm{V-A}}\left( \overline{%
c}_{j}b_{i}\right) _{\mathrm{V-A}},  \label{eq:Operators}
\end{eqnarray}%
$i$ and $j$ are the color indices, and $\left( \overline{q}%
_{1}q_{2}\right) _{\mathrm{V-A}}$ means
\begin{equation}
\left( \overline{q}_{1}q_{2}\right) _{\mathrm{V-A}}=\overline{q}_{1}\gamma
_{\mu }(1-\gamma _{5})q_{2}.  \label{eq:Not}
\end{equation}%
The short-distance Wilson coefficients $c_{1}(\mu )$ and $c_{2}(\mu )$ are
given on the factorization scale $\mu $.

In the factorization method, the amplitude of the decay $T_{b:\overline{s}}^{%
\mathrm{AV}}\rightarrow \mathcal{Z}_{b:\overline{s}}^{0}\rho ^{-}$ has the
form
\begin{eqnarray}
\mathcal{A} &=&\frac{G_{F}}{\sqrt{2}}V_{bc}V_{ud}^{\ast }a(\mu )\langle \rho
^{-}(q)|\left( \overline{d}_{i}u_{i}\right) _{\mathrm{V-A}}|0\rangle  \notag
\\
&&\times \langle \mathcal{Z}_{b:\overline{s}}^{0}(p^{\prime })|\left(
\overline{c}_{j}b_{j}\right) _{\mathrm{V-A}}|T_{b:\overline{s}}^{\mathrm{AV}%
}(p)\rangle ,  \label{eq:Amplitude}
\end{eqnarray}%
where
\begin{equation}
a(\mu )=c_{1}(\mu )+\frac{1}{N_{c}}c_{2}(\mu ),
\end{equation}%
with $N_{c}=3$ being the number of quark colors. The only unknown matrix
element $\langle \rho ^{-}(q)|\left( \overline{d}_{i}u_{i}\right) _{\mathrm{%
V-A}}|0\rangle $ in $\mathcal{A}$ can be defined in the following form

\begin{equation}
\langle \rho ^{-}(q)|\left( \overline{d}_{i}u_{i}\right) _{\mathrm{V-A}%
}|0\rangle =f_{\rho }m_{\rho }\epsilon _{\mu }^{\ast }(q),  \label{eq:ME4}
\end{equation}%
Then, it is evident that
\begin{eqnarray}
\mathcal{A} &=&i\frac{G_{F}}{\sqrt{2}}f_{\rho }V_{bc}V_{ud}^{\ast }a(\mu )%
\left[ \widetilde{G}_{0}(q^{2})\epsilon _{\mu }(p)\epsilon ^{\ast \mu
}(q)\right.  \notag \\
&&+2\widetilde{G}_{1}(q^{2})(p^{\prime }\epsilon (p))(p^{\prime }\epsilon
^{\ast }(q))  \notag \\
&&\left. +i\widetilde{G}_{3}(q^{2})\varepsilon _{\mu \nu \alpha \beta
}\epsilon ^{\ast \mu }(q)\epsilon ^{\nu }(p)p^{\alpha }p^{\prime }{}^{\beta }%
\right] .  \label{eq:Amplitude2}
\end{eqnarray}%
The decay modes $T_{b:\overline{s}}^{\mathrm{AV}}\rightarrow \mathcal{Z}_{b:%
\overline{s}}^{0}K^{\ast }(892)[D^{\ast }(2010)^{-}$, $D_{s}^{\ast -}]$ can
be analyzed in a similar way. To this end, we have to replace in relevant
expressions the spectroscopic parameters ($m_{\rho },f_{\rho }$) of the $%
\rho $ meson with masses and decay constants of the mesons $K^{\ast }(892)$, $%
D^{\ast }(2010)^{-}$, and $D_{s}^{\ast -}$ and make the substitutions $%
V_{ud}\rightarrow V_{us}$, $V_{cd}$, and $V_{cs}$.

The width of the nonleptonic decay $T_{b:\overline{s}}^{\mathrm{AV}%
}\rightarrow \mathcal{Z}_{b:\overline{s}}^{0}\rho ^{-}$ can be evaluated
using the expression
\begin{equation}
\Gamma =\frac{|\mathcal{A}|^{2}}{24\pi m_{\mathrm{AV}}^{2}}\lambda \left( m_{%
\mathrm{AV}},m_{\mathcal{Z}},m_{\rho }\right) ,  \label{eq:NLDW}
\end{equation}%
where
\begin{eqnarray}
\lambda (a,b,c) &=&\frac{1}{2a}\left[ a^{4}+b^{4}+c^{4}\right.  \notag \\
&&\left. -2\left( a^{2}b^{2}+a^{2}c^{2}+b^{2}c^{2}\right) \right] ^{1/2}.
\label{eq:lambda}
\end{eqnarray}%
The key component in Eq.\ (\ref{eq:NLDW}), i.e., $|\mathcal{A}|^{2}$ has a
simple form
\begin{equation}
|\mathcal{A}|^{2}=\sum_{j=0,1,2}H_{j}\widetilde{G}_{j}^{2}+H_{3}\widetilde{G}%
_{0}\widetilde{G}_{1},  \label{eq:A}
\end{equation}%
where $H_{j}$ are given by the expressions%
\begin{eqnarray}
&&H_{0}=\frac{m_{\rho }^{4}+(m_{\mathrm{AV}}^{2}-m_{\mathcal{Z}%
}^{2})^{2}+2m_{\rho }^{2}\left( 5m_{\mathrm{AV}}^{2}-m_{\mathcal{Z}%
}^{2}\right) }{4m_{\rho }^{2}m_{\mathrm{AV}}^{2}},  \notag \\
&&H_{1}=\frac{\left[ m_{\rho }^{4}+(m_{\mathrm{AV}}^{2}-m_{\mathcal{Z}%
}^{2})^{2}-2m_{\rho }^{2}\left( m_{\mathrm{AV}}^{2}+m_{\mathcal{Z}%
}^{2}\right) \right] ^{2}}{4m_{\rho }^{2}m_{\mathrm{AV}}^{2}},  \notag \\
&&H_{2}=\frac{1}{2}\left[ m_{\rho }^{4}+(m_{\mathrm{AV}}^{2}-m_{\mathcal{Z}%
}^{2})^{2}-2m_{\rho }^{2}\left( m_{\mathrm{AV}}^{2}+m_{\mathcal{Z}%
}^{2}\right) \right] ,  \notag \\
&&H_{3}=-\frac{1}{2m_{\rho }^{2}m_{\mathrm{AV}}^{2}}\left[ m_{\rho }^{6}+(m_{%
\mathrm{AV}}^{2}-m_{\mathcal{Z}}^{2})^{3}-m_{\rho }^{4}\left( m_{\mathrm{AV}%
}^{2}\right. \right.  \notag \\
&&\left. \left. +3m_{\mathcal{Z}}^{2}\right) -m_{\rho }^{2}\left( m_{\mathrm{%
AV}}^{4}+2m_{\mathcal{Z}}^{2}m_{\mathrm{AV}}^{2}-3m_{\mathcal{Z}}^{2}\right) %
\right] .  \label{eq:AA}
\end{eqnarray}%
In Eq. \ (\ref{eq:A}), we take into account that the weak form factors $%
\widetilde{G}_{j}$ are real functions of $q^{2}$, and their values for the
process $T_{b:\overline{s}}^{\mathrm{AV}}\rightarrow \mathcal{Z}_{b:%
\overline{s}}^{0}M$ are fixed at $q^{2}=m_{M}^{2}$. \ \

\begin{table}[tbp]
\begin{tabular}{|c|c|}
\hline\hline
Quantity & Value \\ \hline
$m_{\rho} $ & $(775.26 \pm 0.25)~\mathrm{MeV}$ \\
$m_{K^{\star}}$ & $(891.66\pm 0.26)~\mathrm{MeV}$ \\
$m_{D^{\star}}$ & $(2010.26 \pm 0.05)~\mathrm{MeV}$ \\
$m_{D_s^{\star}}$ & $(2112.2 \pm 0.4)~\mathrm{MeV}$ \\
$f_{\rho }$ & $(210 \pm 4)~\mathrm{MeV}$ \\
$f_{K^{\star}}$ & $(204 \pm 7)~\mathrm{MeV}$ \\
$f_{D^{\star}}$ & $(223.5 \pm 8.4)~\mathrm{MeV}$ \\
$f_{D_s^{\star}}$ & $(268.8 \pm 6.6)~\mathrm{MeV}$ \\
$|V_{ud}|$ & $0.97420\pm 0.00021$ \\
$|V_{us}|$ & $0.2243\pm 0.0005$ \\
$|V_{cd}|$ & $0.218\pm 0.004$ \\
$|V_{cs}|$ & $0.997\pm 0.017$ \\ \hline\hline
\end{tabular}%
\caption{Masses and decay constants of the final-state vector mesons and
CKM matrix elements.}
\label{tab:MesonPar}
\end{table}

All input information necessary for numerical analysis is presented in
Table \ref{tab:MesonPar}: the table lists the spectroscopic parameters of the
final-state mesons and the CKM matrix elements. For the masses of the vector
mesons, we use information from PDG \cite{Tanabashi:2018oca}. The decay
constants of mesons $\rho $ and $K^{\ast }(892)$ are also taken from
this source. The decay constants of mesons $D^{\ast }$ and $D_{s}^{\ast }$
are theoretical predictions obtained in the lattice QCD framework
\cite{Lubicz:2016bbi}. The coefficients $c_{1}(m_{b})$, and $c_{2}(m_{b})$
with next-to-leading order QCD corrections have been borrowed from Refs.\
\cite{Buras:1992zv,Ciuchini:1993vr,Buchalla:1995vs}
\begin{equation}
c_{1}(m_{b})=1.117,\ c_{2}(m_{b})=-0.257.  \label{eq:WCoeff}
\end{equation}

For the decay $T_{b:\overline{s}}^{\mathrm{AV}}\rightarrow \mathcal{Z}_{b:%
\overline{s}}^{0}\rho ^{-}$, our calculations yield
\begin{eqnarray}
\Gamma (T_{b:\overline{s}}^{\mathrm{AV}} &\rightarrow &\mathcal{Z}_{b:%
\overline{s}}^{0}\rho ^{-})=\left( 3.47\pm 0.92\right) \times 10^{-10}~%
\mathrm{MeV}.  \notag \\
&&  \label{eq:NLDW1}
\end{eqnarray}%
The partial widths of the remaining three nonleptonic decays are presented below%
\begin{eqnarray}
&&\Gamma (T_{b:\overline{s}}^{\mathrm{AV}}\rightarrow \mathcal{Z}_{b:%
\overline{s}}^{0}K^{\ast }(892))=\left( 1.47\pm 0.37\right) \times 10^{-11}~%
\mathrm{MeV},  \notag \\
&&\Gamma (T_{b:\overline{s}}^{\mathrm{AV}}\rightarrow \mathcal{Z}_{b:%
\overline{s}}^{0}D^{\ast }(2010)^{-})=\left( 1.54\pm 0.39\right) \times
10^{-11}~\mathrm{MeV},  \notag \\
&&\Gamma (T_{b:\overline{s}}^{\mathrm{AV}}\rightarrow \mathcal{Z}_{b:%
\overline{s}}^{0}D_{s}^{\ast }{}^{-})=\left( 4.97\pm 1.32\right) \times
10^{-10}~\mathrm{MeV}.  \notag \\
&&  \label{eq:NLDW2}
\end{eqnarray}%
\newline
It is evident that the parameters of the processes $T_{b:\overline{s}}^{\mathrm{%
AV}}\rightarrow \mathcal{Z}_{b:\overline{s}}^{0}\rho ^{-}$ and $T_{b:%
\overline{s}}^{\mathrm{AV}}\rightarrow \mathcal{Z}_{b:\overline{s}%
}^{0}D_{s}^{\ast }{}^{-}$ are comparable to each other and may affect
predictions for the tetraquark $T_{b:\overline{s}}^{\mathrm{AV}}$: the other two
decays can be safely neglected in the computation of $\Gamma _{\mathrm{full}}$
and $\tau $. Then, using Eqs.\ (\ref{eq:Results}), (\ref{eq:NLDW1}), and (%
\ref{eq:NLDW2}), we find
\begin{eqnarray}
\Gamma _{\mathrm{full}} &=&(12.9\pm 2.1)\times 10^{-8}~\mathrm{MeV},  \notag
\\
\tau &=&5.1_{-0.71}^{+0.99}\times 10^{-15}~\mathrm{s},  \label{eq:WL1}
\end{eqnarray}%
which are principally new predictions of the present article.


\section{Discussion and concluding notes}

\label{sec:Disc}

We have calculated the mass, width, and lifetime of the stable axial-vector
tetraquark $T_{b:\overline{s}}^{\mathrm{AV}}$ with the content $bb\overline{u%
}\overline{s}$. This particle is a strange partner of the tetraquark $%
T_{bb}^{-}$, which was explored in Ref. \cite{Agaev:2018khe}. The width and
lifetime of $T_{bb}^{-}$%
\begin{eqnarray}
\widetilde{\Gamma }_{\mathrm{full}} &=&(7.17\pm 1.23)\times 10^{-8}~\mathrm{%
MeV},  \notag \\
\widetilde{\tau } &=&9.18_{-1.34}^{+1.90}\times 10^{-15}~\mathrm{s},
\label{eq:WL2}
\end{eqnarray}%
are comparable to those of the tetraquark $T_{b:\overline{s}}^{\mathrm{AV}}$%
.

The tetraquark $T_{b:\overline{s}}^{\mathrm{AV}}$ is the last of the four scalar and
axial-vector states $bb\overline{u}\overline{s}$ and $bb\overline{u}%
\overline{d}$ considered in our works. The spectroscopic parameters and
widths of the scalar tetraquarks $T_{b:\overline{s}}^{-}$ and $T_{b:\overline{d}%
}^{-}$ were calculated in Refs.\ \cite{Agaev:2019lwh,Agaev:2020dba}. We
demonstrated there that $T_{b:\overline{s}}^{-}$ and $T_{b:\overline{d}%
}^{-} $ are stable against the strong and electromagnetic decays, and using
the dominant semileptonic and nonleptonic decay channels of these particles,
we estimated their full widths and lifetimes. The information about the
tetraquarks composed of a heavy diquark $bb$ and light antidiquarks is
presented in Table\ \ref{tab:TetraPar}.

It is seen that the scalar particles are heavier than their axial-vector
counterparts: This mass difference for tetraquarks $bb\overline{u}\overline{s}$ is
equal to $35~\mathrm{MeV}$, and for particles with quark content $bb%
\overline{u}\overline{d}$, it reaches $100~\mathrm{MeV}$. It is also clear that
the mass splitting of the strange and nonstrange axial-vector tetraquarks, $%
180~\mathrm{MeV}$, exceeds the value of the same parameter for the scalars, $115~\mathrm{MeV%
}$. These estimates are obtained using the central values of various
tetraquarks' masses calculated using the QCD sum rule method. It
is known that this method is prone to theoretical
uncertainties; therefore, mass splitting between double-beauty tetraquarks
and hierarchy of the particles outlined here must be\ considered with some caution.
Nevertheless, we hope that the picture described above is a quite reliable image
of the real situation.

\begin{widetext}

\begin{table}[tbp]
\begin{tabular}{|c|c|c|c|}
\hline\hline
Tetraquark $(J^{P})$ & Mass ($\mathrm{MeV}$) & Width ($\mathrm{MeV}$) &
Lifetime  \\ \hline
$T_{b:\overline{s}}^{\mathrm{AV}}(1^{+})$ & $10215 \pm 250$ & $(12.9 \pm
2.1)\times 10^{-8}$ & $5.1_{-0.71}^{+0.99}~\mathrm{fs}$  \\
$T_{bb}^{-}(1^{+})$ & $10035 \pm 260$ & $(7.17 \pm 1.23)\times 10^{-8}$ & $%
9.18_{-1.34}^{+1.90}~\mathrm{fs}$   \\
$T_{b:\overline{s}}^{-}(0^{+})$ & $10250 \pm 270$ & $(15.21 \pm 2.59)\times
10^{-10}$ & $0.433_{-0.063}^{+0.089}~\mathrm{ps}$  \\
$T_{b:\overline{d}}^{-}(0^{+})$ & $10135 \pm 240$ & $(10.80 \pm 1.88)\times
10^{-10}$ & $0.605_{-0.089}^{+0.126}~\mathrm{ps}$   \\ \hline\hline
\end{tabular}%
\caption{Parameters of the scalar and axial-vector tetraquarks composed of
the diquark $bb$ and light antidiquarks.}
\label{tab:TetraPar}
\end{table}

\end{widetext}

The widths and lifetimes of these tetraquarks have yielded important
insights into their dynamical properties. It is worth noting that 
the semileptonic decay channels crucially affect the full widths of these tetraquarks: 
our investigations have shown that the partial width of the semileptonic
decay is enhanced relative to the nonleptonic one by $2-3$ orders of magnitude.
The widths of the scalar tetraquarks $T_{b:\overline{s}}^{-}$ and $T_{b:%
\overline{d}}^{-}$ are considerably smaller than the widths of the axial-vector
particles $T_{bb}^{-}$ and $T_{b:\overline{s}}^{\mathrm{AV}}$. As a result,
the mean lifetimes of the scalar tetraquarks are $\sim1\mathrm{ps}$,
whereas for the axial vector states, we get $\tau \approx 10\ \mathrm{fs}$.
Stated differently, the scalar tetraquarks $T_{b:\overline{s}}^{-}$ and $T_{b:%
\overline{d}}^{-}$ are heavier and live longer than the corresponding
axial-vector particles.

The spectroscopic parameters and lifetimes of the axial-vector states $%
T_{bb}^{-}$ and $T_{b:\overline{s}}^{\mathrm{AV}}$ were also explored in
Refs. \ \cite{Karliner:2017qjm,Ali:2018ifm}. The lifetime $367~\mathrm{fs}$
of the state $T_{bb\text{ }}^{-}$ predicted in Ref. \cite{Karliner:2017qjm}
is considerably longer than our result $9.18~\mathrm{fs.}$ The lifetimes $%
\tau \simeq 800~\mathrm{fs}$ of the tetraquarks $T_{bb}^{-}$ and $T_{b:%
\overline{s}}^{\mathrm{AV}}$ obtained in Ref. \cite{Ali:2018ifm} exceed our
predictions as well. Let us note that, in Ref.\ \cite{Ali:2018ifm}, the
authors considered only nonleptonic decays of the axial-vector tetraquarks.
We have reevaluated the lifetime of $T_{b:\overline{s}}^{\mathrm{AV}}$ using
Eqs.\ (\ref{eq:NLDW1}) and (\ref{eq:NLDW2}) and found $\tau \simeq 753~%
\mathrm{fs}$. Despite the fact that the channels that were explored in Ref.\ \cite%
{Ali:2018ifm} differ from the decays that were considered in the present work, for $\tau $,
they lead to compatible predictions. One of the reasons is that, in both cases,
the amplitudes of the nonleptonic weak decays contain two CKM matrix elements, which
suppress their partial widths and branching ratios relative to the semileptonic
channels. Evidently, our results for the nonleptonic decays of $T_{b:\overline{s}%
}^{\mathrm{AV}}$ can be refined by including into analysis some relevant channels
from Ref.\ \cite{Ali:2018ifm}. However, for  discovering stable exotic
mesons, their semileptonic decays seem to be more promising than other
processes.

\section*{ACKNOWLEDGEMENTS}

The work of K.~A, B.~B., and H.~S was supported in part by the TUBITAK grant
under No: 119F050.

\appendix*

\section{ The propagators $S_{q(Q)}(x)$ and invariant amplitude $\Pi
(M^{2},s_{0})$}

\renewcommand{\theequation}{\Alph{section}.\arabic{equation}} \label{sec:App}
\begin{widetext}

In the present work, we use the light quark propagator $S_{q}^{ab}(x)$, which
is given by the following formula
\begin{eqnarray}
&&S_{q}^{ab}(x)=i\delta _{ab}\frac{\slashed x}{2\pi ^{2}x^{4}}-\delta _{ab}%
\frac{m_{q}}{4\pi ^{2}x^{2}}-\delta _{ab}\frac{\langle \overline{q}q\rangle
}{12}+i\delta _{ab}\frac{\slashed xm_{q}\langle \overline{q}q\rangle }{48}%
-\delta _{ab}\frac{x^{2}}{192}\langle \overline{q}g_{s}\sigma Gq\rangle
\notag \\
&&+i\delta _{ab}\frac{x^{2}\slashed xm_{q}}{1152}\langle \overline{q}%
g_{s}\sigma Gq\rangle -i\frac{g_{s}G_{ab}^{\alpha \beta }}{32\pi ^{2}x^{2}}%
\left[ \slashed x{\sigma _{\alpha \beta }+\sigma _{\alpha \beta }}\slashed x%
\right] -i\delta _{ab}\frac{x^{2}\slashed xg_{s}^{2}\langle \overline{q}%
q\rangle ^{2}}{7776}  \notag \\
&&-\delta _{ab}\frac{x^{4}\langle \overline{q}q\rangle \langle
g_{s}^{2}G^{2}\rangle }{27648}+\cdots .
\end{eqnarray}%
For the heavy quarks $Q$, we utilize the propagator $S_{Q}^{ab}(x)$
\begin{eqnarray}
&&S_{Q}^{ab}(x)=i\int \frac{d^{4}k}{(2\pi )^{4}}e^{-ikx}\Bigg \{\frac{\delta
_{ab}\left( {\slashed k}+m_{Q}\right) }{k^{2}-m_{Q}^{2}}-\frac{%
g_{s}G_{ab}^{\alpha \beta }}{4}\frac{\sigma _{\alpha \beta }\left( {\slashed %
k}+m_{Q}\right) +\left( {\slashed k}+m_{Q}\right) \sigma _{\alpha \beta }}{%
(k^{2}-m_{Q}^{2})^{2}}  \notag \\
&&+\frac{g_{s}^{2}G^{2}}{12}\delta _{ab}m_{Q}\frac{k^{2}+m_{Q}{\slashed k}}{%
(k^{2}-m_{Q}^{2})^{4}}+\frac{g_{s}^{3}G^{3}}{48}\delta _{ab}\frac{\left( {%
\slashed k}+m_{Q}\right) }{(k^{2}-m_{Q}^{2})^{6}}\left[ {\slashed k}\left(
k^{2}-3m_{Q}^{2}\right) +2m_{Q}\left( 2k^{2}-m_{Q}^{2}\right) \right] \left(
{\slashed k}+m_{Q}\right) +\cdots \Bigg \}.  \notag \\
&&
\end{eqnarray}

Above, we have used the notation
\begin{equation}
G_{ab}^{\alpha \beta }\equiv G_{A}^{\alpha \beta }t_{ab}^{A},\ \
\end{equation}%
where $G_{A}^{\alpha \beta }$ is the gluon field strength tensor, and $%
t^{A}=\lambda ^{A}/2$ with $\lambda ^{A}$ being the Gell-Mann matrices, $%
A=1,2,\cdots ,8.$

The invariant amplitude $\Pi ^{\mathrm{OPE}}(p^{2})$ used for calculating
the mass and coupling of the tetraquark $T_{b:\overline{s}}^{-}$ after the
Borel transformation and subtraction procedures takes the following form%
\begin{equation}
\Pi (M^{2},s_{0})=\int_{\mathcal{M}^{2}}^{s_{0}}ds\rho ^{\mathrm{OPE}%
}(s)e^{-s/M^{2}}+\Pi (M^{2}),
\end{equation}%
where
\begin{equation}
\rho ^{\mathrm{OPE}}(s)=\rho ^{\mathrm{pert.}}(s)+\sum_{N=3}^{8}\rho ^{%
\mathrm{DimN}}(s),\ \ \Pi (M^{2})=\sum_{N=6}^{10}\Pi ^{\mathrm{DimN}}(M^{2}).
\label{eq:A1}
\end{equation}%
Components of the spectral density are given by the formulas%
\begin{equation}
\rho (s)=\int_{0}^{1}d\alpha \int_{0}^{1-a}d\beta \rho (s,\alpha ,\beta ),\
\ \rho (s)=\int_{0}^{1}d\alpha \rho (s,\alpha ),  \label{eq:A2}
\end{equation}%
depending on whether $\rho (s,\alpha ,\beta )$ is a function of $\alpha $
and $\beta $ or only $\alpha$. The same is true also for terms $\Pi (M^{2})$%
, i.e.,%
\begin{equation}
\Pi ^{\mathrm{DimN}}(M^{2})=\int_{0}^{1}d\alpha \int_{0}^{1-a}d\beta \Pi ^{%
\mathrm{DimN}}(M^{2},\alpha ,\beta ),\ \ \Pi (M^{2})=\int_{0}^{1}d\alpha \Pi
^{\mathrm{DimN}}(M^{2},\alpha ).  \label{eq:A4}
\end{equation}%
In these expressions, $\alpha $ and $\beta $ are Feynman parameters.

The perturbative and nonperturbative contributions of dimensions $3$, $4$,
and $5$ are terms of (\ref{eq:A2}) types. For relevant spectral densities,
we get%
\begin{equation}
\rho ^{\mathrm{pert.}}(s,\alpha ,\beta )=\frac{\Theta (L_{1})}{2048\pi
^{6}L^{2}N_{1}^{7}}\left[ s\alpha \beta L-m_{b}^{2}N_{2}\right] ^{3}\left\{
5s\alpha \beta L^{2}+m_{b}^{2}N_{1}\left[ 3\beta ^{2}+3\alpha (\alpha
-1)+\beta \left( 2\alpha -3\right) \right] \right\} ,
\end{equation}%
\begin{eqnarray}
&&\rho ^{\mathrm{Dim3}}(s,\alpha ,\beta )=\frac{m_{s}\left[ 2\langle
\overline{u}u\rangle -\langle \overline{s}s\rangle \right] }{128\pi
^{4}N_{1}^{5}}\Theta (L_{1})\left\{ -3s^{2}\alpha ^{2}\beta
^{2}L^{3}+m_{b}^{4}(\alpha +\beta )N_{1}^{2}[\alpha (\alpha -1)+\beta (\beta
-1)]+2m_{b}^{2}s\alpha \beta \right.   \notag \\
&&\left. \times \left[ \beta ^{5}+\alpha ^{2}(\alpha -1)^{3}+\beta
^{4}(5\alpha -3)+\alpha \beta (\alpha -1)^{2}(5\alpha -2)+3\beta
^{3}(1-4\alpha +3\alpha ^{2})-\beta ^{2}(1-9\alpha +17\alpha ^{2}-9\alpha
^{3})\right] \right\} ,  \notag \\
&&
\end{eqnarray}%
\begin{eqnarray}
&&\rho ^{\mathrm{Dim4}}(s,\alpha ,\beta )=\frac{\langle \alpha _{s}G^{2}/\pi
\rangle }{6144\pi ^{4}(1-\beta )L^{2}N_{1}^{5}}\Theta (L_{1})\left\{
-s^{2}\alpha ^{2}\beta ^{2}(\beta -1)L^{3}\left[ 18\beta ^{2}+18(\alpha
-1)^{2}+\beta (31\alpha -36)\right] \right.   \notag \\
&&+m_{b}^{4}N_{1}^{2}\left[ 10\beta ^{6}+\beta ^{5}(21\alpha -32)+\beta
^{4}(40-76\alpha +29\alpha ^{2}+\beta ^{3}(-24+97\alpha -95\alpha
^{2}+37\alpha ^{3})\right.   \notag \\
&&+2\alpha ^{2}(3-9\alpha +11\alpha ^{2}-9\alpha ^{3}+4\alpha ^{4})+\beta
\alpha (12-48\alpha +73\alpha ^{2}-59\alpha ^{3}+26\alpha ^{4})  \notag \\
&&\left. +\beta ^{2}(6-54\alpha +108\alpha ^{2}-92\alpha ^{3}+37\alpha ^{4})
\right] +4sm_{b}^{2}\alpha \beta LN_{1}\left[ 3\beta (\beta -1)^{4}+\alpha
(\beta -1)(-3+21\beta -32\beta ^{2}+16\beta ^{3})\right.   \notag \\
&&\left. \left. +\alpha ^{2}(\beta -1)(9-32\beta +22\beta ^{2})+\alpha
^{3}(\beta -1)(-9+14\beta )+\alpha ^{4}(\beta -1)-2\alpha ^{5}\right]
\right\} ,
\end{eqnarray}%
\begin{equation}
\rho ^{\mathrm{Dim5}}(s,\alpha )=\frac{m_{s}\left[ 3\langle \overline{u}%
g_{s}\sigma Gu\rangle -\langle \overline{s}g_{s}\sigma Gs\rangle \right] }{%
384\pi ^{4}}\Theta (L_{2})(2m_{b}^{2}+s-3s\alpha +2s\alpha ^{2}).
\end{equation}%
The $\mathrm{DimN}=6$, $7$ and $8$ terms have mixed compositions: they
contain components expressed through both $\rho ^{\mathrm{DimN}}(s)$ and $%
\Pi ^{\mathrm{DimN}}(M^{2})$. For these terms, we find
\begin{eqnarray}
\Pi ^{\mathrm{DimN}}(M^{2},s_{0}) &=&\int_{\mathcal{M}%
^{2}}^{s_{0}}dse^{-s/M^{2}}\int_{0}^{1}d\alpha \int_{0}^{1-a}d\beta \rho
_{1}^{\mathrm{DimN}}(s,\alpha ,\beta )+\int_{\mathcal{M}%
^{2}}^{s_{0}}dse^{-s/M^{2}}\int_{0}^{1}d\alpha \rho _{2}^{\mathrm{DimN}%
}(s,\alpha )  \notag \\
&&+\int_{0}^{1}d\alpha \int_{0}^{1-a}d\beta \Pi ^{\mathrm{DimN}%
}(M^{2},\alpha ,\beta ).  \label{eq:A5}
\end{eqnarray}%
In the case of $\mathrm{DimN}=6$, the relevant functions have the expressions%
\begin{equation}
\rho _{1}^{\mathrm{Dim6}}(s,\alpha ,\beta )=-\frac{\langle
g_{s}^{3}G^{3}\rangle m_{b}^{2}\alpha ^{5}}{10240\pi ^{6}(\beta -1)LN_{1}^{3}%
}\Theta (L_{1}),\
\end{equation}%
\begin{equation}
\rho _{2}^{\mathrm{Dim6}}(s,\alpha )=\frac{\Theta (L_{2})}{24\pi ^{2}}\left[
\langle \overline{s}s\rangle \langle \overline{u}u\rangle +\frac{g_{s}^{2}}{%
108\pi ^{2}}(\langle \overline{s}s\rangle ^{2}+\langle \overline{u}u\rangle
^{2})\right] (2m_{b}^{2}+s-3s\alpha +2s\alpha ^{2}),
\end{equation}%
\begin{eqnarray}
&&\Pi ^{\mathrm{Dim6}}(M^{2},\alpha ,\beta )=-\frac{\langle
g_{s}^{3}G^{3}\rangle m_{b}^{4}}{30720M^{2}\pi ^{6}\alpha ^{2}\beta
^{2}L^{4}N_{1}^{3}}\exp \left[ -\frac{m_{b}^{2}}{M^{2}}\frac{N_{1}(\alpha
+\beta )}{\alpha \beta L}\right]   \notag \\
&&\times \left\{ m_{b}^{2}(\alpha +\beta )N_{1}\left[ 5\beta ^{8}+2\beta
^{5}\alpha ^{2}(3-4\alpha )+2\beta ^{3}\alpha ^{4}(5-4\alpha )+3\alpha
^{6}\beta (\alpha -1)+5\alpha ^{6}(\alpha -1)^{2}\right. \right.   \notag \\
&&\left. +\beta ^{7}(-10+3\alpha )+\beta ^{4}\alpha ^{2}(-5+2\alpha
(5-4\alpha ))-\beta ^{2}\alpha ^{4}(5+\alpha (-6+\alpha ))-\beta
^{6}(-5+\alpha (3+\alpha ))\right]   \notag \\
&&+M^{2}\alpha \beta L\left[ 11\beta ^{8}+8\beta ^{3}\alpha ^{4}+11\alpha
^{6}(\alpha -1)^{2}+3\beta \alpha ^{5}(\alpha -1)(-5+6\alpha )+\beta
^{5}\alpha (\alpha -1)(-15+8\alpha )\right.   \notag \\
&&\left. \left. +2\beta ^{7}(-11+9\alpha )+\beta ^{2}\alpha ^{4}(\alpha
-1)(-4+19\alpha )+4\alpha ^{2}\beta ^{4}(1+2\alpha (\alpha -1))+\beta
^{6}(11+\alpha (-33+19\alpha )\right] \right\} .
\end{eqnarray}%
Contribution of dimension $7$ is determined by the same formula (\ref%
{eq:A5}), where $\rho _{1}^{\mathrm{Dim7}}(s,\alpha ,\beta )$, $\rho _{2}^{%
\mathrm{Dim7}}(s,\alpha )$ and $\Pi ^{\mathrm{Dim7}}(M^{2},\alpha ,\beta )$
are given by the following expressions:
\begin{equation}
\rho _{1}^{\mathrm{Dim7}}(s,\alpha ,\beta )=\frac{\langle \alpha
_{s}G^{2}/\pi \rangle m_{s}\left[ 2\langle \overline{u}u\rangle -\langle
\overline{s}s\rangle \right] }{768\pi ^{2}N_{1}^{3}}\Theta (L_{1})\alpha
\beta L,\ \rho _{2}^{\mathrm{Dim7}}(s,\alpha )=-\frac{\langle \alpha
_{s}G^{2}/\pi \rangle m_{s}\langle \overline{u}u\rangle }{1152\pi ^{2}}%
\Theta (L_{2})(1-4\alpha +3\alpha ^{2}),
\end{equation}%
\begin{eqnarray}
&&\Pi ^{\mathrm{Dim7}}(M^{2},\alpha ,\beta )\ =\frac{\langle \alpha
_{s}G^{2}/\pi \rangle m_{b}^{2}m_{s}\left[ \langle \overline{s}s\rangle
-2\langle \overline{u}u\rangle \right] }{2304M^{2}\pi ^{2}\alpha ^{2}\beta
^{2}(\beta -1)LN_{1}^{3}}\exp \left[ -\frac{m_{b}^{2}}{M^{2}}\frac{%
N_{1}(\alpha +\beta )}{\alpha \beta L}\right] \left\{ 2m_{b}^{2}(\beta
-1)(\alpha +\beta )^{2}\right.   \notag \\
&&\times \left[ \beta ^{4}+\beta ^{3}(\alpha -1)+\beta \alpha ^{2}(\alpha
-1)+\alpha ^{3}(\alpha -1)+\beta ^{2}\alpha (2\alpha -1)\right] -M^{2}\alpha
\beta   \notag \\
&&\left. \times \left[ 4\beta ^{5}+\beta ^{4}(\alpha -8)+4\alpha ^{3}(\alpha
-1)^{2}+2\beta ^{2}(2-\alpha +\alpha ^{2})+\beta ^{2}\alpha (1-3\alpha
+5\alpha ^{2})+\beta \alpha ^{2}(1-9\alpha +8\alpha ^{2})\right] \right\} .
\end{eqnarray}

The relevant functions for dimension $8$ are%
\begin{eqnarray}
&&\rho _{1}^{\mathrm{Dim8}}(s,\alpha ,\beta )=-\frac{\langle \alpha
_{s}G^{2}/\pi \rangle ^{2}}{6144\pi ^{2}N_{1}^{3}}\Theta (L_{1})\alpha \beta
(\alpha +\beta -1),\ \ \rho _{2}^{\mathrm{Dim8}}(s,\alpha )=-\frac{\langle
\overline{s}g_{s}\sigma Gs\rangle \langle \overline{u}u\rangle }{48\pi ^{2}}%
\Theta (L_{2})\ (1-4\alpha +3\alpha ^{2}),  \notag \\
&&\Pi ^{\mathrm{Dim8}}(M^{2},\alpha ,\beta )\ =-\frac{\langle \alpha
_{s}G^{2}/\pi \rangle ^{2}m_{b}^{2}}{27648M^{4}\pi ^{2}\alpha ^{2}\beta
^{2}(\beta -1)L^{4}N_{1}^{3}}\exp \left[ -\frac{m_{b}^{2}}{M^{2}}\frac{%
N_{1}(\alpha +\beta )}{\alpha \beta L}\right] \left\{ m_{b}^{4}\alpha
^{2}\beta ^{2}(\alpha +\beta )(\beta -1)N_{1}^{2}\right.  \notag \\
&&\times \left[ 2\beta ^{2}+2\alpha (\alpha -1)+\beta (3\alpha -2)\right]
+M^{4}\alpha \beta L^{2}\left[ 6\beta ^{8}+6\alpha ^{4}(\alpha
-1)^{4}+3\beta ^{7}(5\alpha -8)+3\alpha ^{3}\beta (\alpha -1)^{3}(8\alpha
-3)\right.  \notag \\
&&+\beta ^{6}(36-54\alpha +26\alpha ^{2})+\beta ^{2}\alpha ^{2}(\alpha
-1)^{2}(6-39\alpha +47\alpha ^{2})+\beta ^{5}(-24+72\alpha -82\alpha
^{2}+33\alpha ^{3})  \notag \\
&&\left. +\beta ^{4}(6-42\alpha +92\alpha ^{2}-99\alpha ^{3}+47\alpha
^{4})+\alpha \beta ^{3}(9-42\alpha +108\alpha ^{2}-133\alpha ^{3}+58\alpha
^{4})\right]  \notag \\
&&-m_{b}^{2}M^{2}LN_{1}\left[ 3\beta ^{5}(\beta -1)^{4}+3\alpha \beta
^{4}(\beta -1)^{3}(-3+5\beta )+\alpha ^{2}\beta ^{3}(\beta -1)^{2}(12+\beta
(-45+38\beta ))\right.  \notag \\
&&+2\alpha ^{3}\beta ^{2}(\beta -1)^{2}(6+\beta (-27+31\beta ))+\alpha
^{4}\beta (\beta -1)(-9+\beta (-57+\beta (-116+73\beta )))  \notag \\
&&+\alpha ^{5}(\beta -1)(-3+\beta (33+\beta (-83+66\beta )))+\alpha
^{6}(-9+\beta (48+\beta (-79+42\beta )))+\alpha ^{7}(9+\beta (-24+17\beta ))
\notag \\
&&\left. \left. +3\alpha ^{8}(\beta -1)\right] \right\} +\frac{\langle
\alpha _{s}G^{2}/\pi \rangle ^{2}m_{b}^{2}(\alpha +\beta )}{18432\pi
^{2}N_{1}^{2}}\exp \left[ -\frac{m_{b}^{2}}{M^{2}}\frac{N_{1}(\alpha +\beta )%
}{\alpha \beta L}\right].
\end{eqnarray}%
The $\mathrm{Dim9}$ and $\mathrm{Dim10}$ contributions are exclusively of the (%
\ref{eq:A4}) types
\begin{equation}
\Pi ^{\mathrm{DimN}}(M^{2},s_{0})=\int_{0}^{1}d\alpha \int_{0}^{1-a}d\beta
\Pi _{1}^{\mathrm{DimN}}(M^{2},\alpha ,\beta )+\int_{0}^{1}d\alpha \Pi _{2}^{%
\mathrm{DimN}}(M^{2},\alpha ).
\end{equation}%
For $\mathrm{Dim9}$, we get%
\begin{equation}
\Pi _{1}^{\mathrm{Dim9}}(M^{2},\alpha ,\beta )=\frac{\langle
g_{s}^{3}G^{3}\rangle m_{b}^{2}m_{s}\left[ 2\langle \overline{u}u\rangle
-\langle \overline{s}s\rangle \right] }{23040M^{6}\pi ^{4}\alpha ^{4}\beta
^{4}(\beta -1)L^{4}N_{1}^{2}}R_{1}(M^{2},\alpha ,\beta ),
\end{equation}%
and%
\begin{equation}
\Pi _{2}^{\mathrm{Dim9}}(M^{2},\alpha )=\frac{\langle \alpha _{s}G^{2}/\pi
\rangle m_{s}\left[ \langle \overline{s}g_{s}\sigma Gs\rangle -3\langle
\overline{u}g_{s}\sigma Gu\rangle \right] }{13824M^{4}\pi ^{2}\alpha
^{4}(\alpha -1)^{2}}R_{2}(M^{2},\alpha ).
\end{equation}%
The dimension $10$ term has the following components:%
\begin{equation}
\Pi _{1}^{\mathrm{Dim10}}(M^{2},\alpha ,\beta )=-\frac{\langle \alpha
_{s}G^{2}/\pi \rangle \langle g_{s}^{3}G^{3}\rangle m_{b}^{2}}{%
184320M^{6}\pi ^{4}\alpha ^{4}\beta ^{4}(\beta -1)L^{4}N_{1}^{2}}%
R_{1}(M^{2},\alpha ,\beta ),
\end{equation}%
and%
\begin{equation}
\Pi _{2}^{\mathrm{Dim10}}(M^{2},\alpha )=-\frac{\langle \alpha _{s}G^{2}/\pi
\rangle }{864M^{4}\alpha ^{4}(\alpha -1)^{2}}\left[ \langle \overline{s}%
s\rangle \langle \overline{u}u\rangle +\frac{g_{s}^{2}}{108\pi ^{2}}(\langle
\overline{s}s\rangle ^{2}+\langle \overline{u}u\rangle ^{2})\right]
R_{2}(M^{2},\alpha),
\end{equation}%
where functions $R_{1}(M^{2},\alpha ,\beta )$ and $R_{2}(M^{2},\alpha )$ are
given by the formulas
\begin{eqnarray}
&&R_{1}(M^{2},\alpha ,\beta )=\exp \left[ -\frac{m_{b}^{2}}{M^{2}}\frac{%
N_{1}(\alpha +\beta )}{\alpha \beta L}\right] \left\{ -2M^{4}\alpha
^{2}\beta ^{2}L^{3}\left[ 3\beta ^{7}+\beta ^{6}(\alpha -6)-\beta ^{4}\alpha
(\alpha -1)+\alpha ^{4}\beta ^{3}+3\alpha ^{5}(\alpha -1)^{2}\right. \right.
\notag \\
&&\left. +2\beta ^{2}\alpha ^{4}(2\alpha -1)+\beta ^{5}(3-2\alpha +\alpha
^{2})+\beta \alpha (1-7\alpha +6\alpha ^{2})\right] +m_{b}^{4}(\beta
-1)N_{1}^{2}\left[ 5\beta ^{9}+5\alpha ^{7}(\alpha -1)^{2}+2\beta
^{8}(-5+4\alpha )\right.  \notag \\
&&\left. +\beta ^{5}\alpha ^{2}(-5+16\alpha -16\alpha ^{2})+\beta \alpha
^{6}(5-13\alpha +8\alpha ^{2})\right] +m_{b}^{2}M^{2}\alpha \beta LN_{1}%
\left[ 5\beta ^{6}(\beta -1)^{3}+3\beta ^{5}\alpha (\beta -1)^{2}(-5+6\beta
)\right.  \notag \\
&&+\beta ^{4}\alpha ^{2}(\beta -1)(L+\alpha )(-16+35\beta )+16\beta
^{2}\alpha ^{4}(\beta -1)(1-2\beta +2\beta ^{2})+\beta \alpha ^{5}(\beta
-1)^{2}(-21+41\beta )  \notag \\
&&\left. \left. +\alpha ^{6}(\beta -1)(5+\beta (-61+60\beta ))+3\alpha
^{7}(\beta -1)(-7+18\beta )+3\alpha ^{8}(-9+10\beta )+11\alpha ^{9}\right]
\right\},
\end{eqnarray}%
and%
\begin{eqnarray}
R_{2}(M^{2},\alpha ) &=&\exp \left[ -\frac{m_{b}^{2}}{M^{2}\alpha (1-\alpha )%
}\right] \left[ M^{4}\alpha ^{3}(\alpha -1)^{2}(1+2\alpha
)-4m_{b}^{4}(1-3\alpha +3\alpha ^{2})\right.  \notag \\
&&\left. +m_{b}^{2}M^{2}\alpha (8-27\alpha +32\alpha ^{2}-7\alpha ^{3})
\right].
\end{eqnarray}

In the expressions above, $\Theta (z)$ is the unit step function. We have used also
the following short-hand notations:%
\begin{eqnarray}
N_{1} &=&\beta ^{2}+\beta (\alpha -1)+\alpha (\alpha -1),\ \ N_{2}=(\alpha
+\beta )N_{1},\ \ L=\alpha +\beta -1,\   \notag \\
L_{1} &=&\frac{(1-\beta )}{N_{1}^{2}}\left[ m_{b}^{2}N_{2}-s\alpha \beta L%
\right] ,\ \ \ \ L_{2}=s\alpha (1-\alpha )-m_{b}^{2}.
\end{eqnarray}

\end{widetext}

\end{document}